\documentclass[12pt]{article}
\usepackage{epsfig}
\usepackage{pst-plot,url,epsf}
\usepackage{t1enc}
\usepackage{placeins}
\usepackage{cite}
\newcommand{\beq}{\begin{equation}}
\newcommand{\eeq}{\end{equation}}
\newcommand{\bea}{\begin{eqnarray}}
\newcommand{\eea}{\end{eqnarray}}

\newcommand{\epm}{e^+e^-}
\newcommand{\nn}{\nonumber}

\def\earr{\end{array}}
\def\barr#1{\begin{array}{#1}}

\begin{document}
\thispagestyle{empty}
\begin{flushright}
3 January, 2017\\
Revised version:\\
15 March, 2017\\
DESY 17-001, HU-EP-17/01\\
\vspace*{1.5cm}
\end{flushright}
\begin{center}
  {\LARGE\bf Photon radiation in $\epm\;\to\;$hadrons at low \\[4mm]
    energies with {\tt carlomat\_3.1}}

\vspace*{1cm}
Fred Jegerlehner\footnote{E-mail: fjeger@physik.hu-berlin.de}\\[1cm]
{\small\it Deutsches Elektronen--Synchrotron (DESY), Platanenallee 6,
15738 Zeuthen, Germany\\
Humboldt--Universit\"at zu Berlin, Institut f\"ur Physik, Newtonstrasse 15,
12489 Berlin, Germany}\\[1.cm]
and\\[1.cm]
Karol Ko\l odziej\footnote{E-mail: karol.kolodziej@us.edu.pl}\\[1cm]
{\small\it
Institute of Physics, University of Silesia\\ 
ul. Uniwersytecka 4, 40007 Katowice, Poland}\\
\vspace*{2.5cm}
{\bf Abstract}\\
\end{center}
We present a sample of results for the cross sections of
several processes of low energetic $\epm$ annihilation into
final states containing pions accompanied by one or two photons, or a light
lepton pair. The results, which have been obtained with a new version
of a multipurpose Monte Carlo program {\tt carlomat}, labelled 3.1,
demonstrate new capabilities of the program which, among others, include a
possibility of taking into account either the initial or final
state radiation separately, or both at a time, and a possibility of inclusion of
the electromagnetic charged pion form factor for processes with charged pion pairs.
We also discuss some problems related to the $U(1)$ electromagnetic gauge
invariance.

\vfill

\newpage

\section{Introduction}

Better determination of the hadronic contribution to vacuum
polarisation is indispensable to increase precision of theoretical
predictions for the muon and electron $g-2$ and to improve evolution
of the fine structure constant from the Thomson limit to high energy
scales. For example, $\alpha(M_Z)$ becomes particularly relevant in
the context of the high energy $\epm$ collider projects
\cite{ILC, ECFA/DESY, ALC, ACFA, CLIC, FCCee, CEPC1, CEPC2}, which are more and more intensively
discussed in the recent years, some of them including the giga-$Z$
option.  The hadronic contribution to vacuum polarisation can be
derived, with the help of dispersion relations, from the energy
dependence of the ratio $R_{\gamma}(s)\equiv \sigma^{(0)}(e^+e^-\to
\gamma^*\to {\rm hadrons})
/\frac{4\pi\alpha^2}{3s}$~\cite{CabibboGatto60, CabibboGatto61}. For early data driven
analyses, see e.g.~\cite{EJ95,ADH98}, and for the sufficiently precise
perturbative QCD results, which are needed for the perturbatively accessible
windows and the high energy tail, see~\cite{KS98} and the
references therein.  Presently one is including corrections to four
loops~\cite{HS02}. For more recent $\alpha(M_Z)$ and $g-2$ evaluations
we refer to~\cite{HMNT11,Zhang:2015yfi,Davier:2016udg}. A recent
update of the hadronic contribution to the electron and muon
$g-2$~\cite{fjeger} also accounts for the current $\epm$ data situation, as
summarised below. In the regions from 5.2 to 9.46~GeV and above
13~GeV, perturbative QCD is applied. One of the main issues is
$R_\gamma(s)$ in the region from 1.2 to 2.0~GeV, where more than
30 exclusive channels must be measured.

In the low energy region, which is particularly important for the dispersive
evaluation of the hadronic contribution to the muon $g-2$, data have improved
dramatically in the past decade for the dominant $e^+e^- \to
\pi^+\pi^-$ channel (CMD-2~\cite{NSK1, NSK2, NSK3, NSK4}, SND/Novosibirsk~\cite{NSK5},
KLOE/Frascati~\cite{KLOE1, KLOE1_1, KLOE2, KLOE3},
BaBar/SLAC~\cite{BaBar, BaBar1},
BES-III/Beijing~\cite{BESIII}) and the statistical errors are a minor
problem now. Similarly the important region between 1.2 to 2.4 GeV
has been improved a lot by the BaBar exclusive channel measurements in the ISR
mode~\cite{BaBar05, BaBar05_1, BaBar05_2, BaBar05_3, BaBar07, BaBar07_2, BaBar07_3, BaBar11, BaBar11_1, BaBar11_2, BaBar11_3, BaBar11_4, Lees:2013ebn, Lees:2013gzt, Lees:2014xsh, Davier:2015bka}.
Recent data sets collected are: $e^+e^-\to 3(\pi^+\pi^-)$, $e^+e^-\to
\bar{p}p$ and $e^+ e^- \to K^0_{S}K^0_{L}$ from
CMD-3~\cite{Akhmetshin:2013xc,Akhmetshin:2015ifg,Kozyrev:2016raz},
and $e^+e^- \to \omega\pi^0 \to \pi^0\pi^0\gamma$, $e^+e^-\to \eta \pi^+\pi^-$
and $e^+e^-\to \pi^0\gamma$ from
SND~\cite{Achasov:2013btb,Aulchenko:2014vkn,Achasov:2016bfr}.
Above 2~GeV fairly accurate BES-II data~\cite{BES02, BES02_1, Ablikim:2009ad} are
available. Recently, a new inclusive determination of $R_\gamma(s)$ in
the range 1.84--3.72 GeV has been obtained with the KEDR detector 
at Novosibirsk~\cite{Anashin:2015woa,Anashin:2016hmv}. However, the
contribution from the range above 1 GeV is still
contributing about 50\% to the hadronic uncertainty of $a_\mu^{\rm had}$.

The main part of the hadronic uncertainty of $a_\mu^{\rm had}$
is related to the systematic errors of the experimental data, 
but the uncertainties due to missing radiative corrections, which are dominated
by photon radiation corrections, are quite relevant too.

To obtain reliable theoretical predictions for that many hadronic processes is 
a challenge indeed. It is obvious that the correct description of the most relevant 
hadronic channels as, e.g., $\pi^+\pi^-$, requires the inclusion of radiative corrections. 
This demand is successfully met by the dedicated Monte Carlo (MC) generator 
{\tt PHOKHARA} \cite{PHOKHARA}.
However, for many sub-dominant channels, with three or more particles in the final state, 
it is probably enough to have the leading order 
(LO) predictions. If those channels are measured with the method of radiative return the
predictions must also include at least the initial state radiation photons.
Production of hadrons at low energies, as well as the photon
radiation off them, is usually described in the framework of some effective model
which often includes quite a number of interaction vertices and mixing terms.
Thus, it is obvious that the number of Feynman diagrams of such sub-dominant
multiparticle processes may become quite big. Therefore,
in order to obtain a reliable description of relatively many potentially interesting 
hadronic processes it is required to fully automate the process of MC code generation.
This requirement has been met by 
version 3.0 of {\tt carlomat}, a program that allows one to generate automatically
the Monte Carlo programs dedicated to the description of, among others, the processes 
$\epm\to{\rm hadrons}$ at low centre-of-mass energies \cite{carlomat3}.
In addition to the standard model (SM) and some of its extensions, {\tt carlomat\_3.0}
includes the Feynman rules of the scalar electrodynamics (sQED), the effective 
vertices of electromagnetic (EM) interaction of spin 1/2 nucleons and a number 
of triple and quartic vertices and mixing terms resulting from the resonance 
chiral theory (R$\chi$T) or hidden local symmetry (HLS) model. Although there are
options in the program that potentially allow one to include momentum dependence
in any of the effective couplings, no such a running coupling, except for the EM
form factors of the nucleons, has been actually implemented in it.

\begin{figure}[h]
\centering
\includegraphics[width=0.45\textwidth]{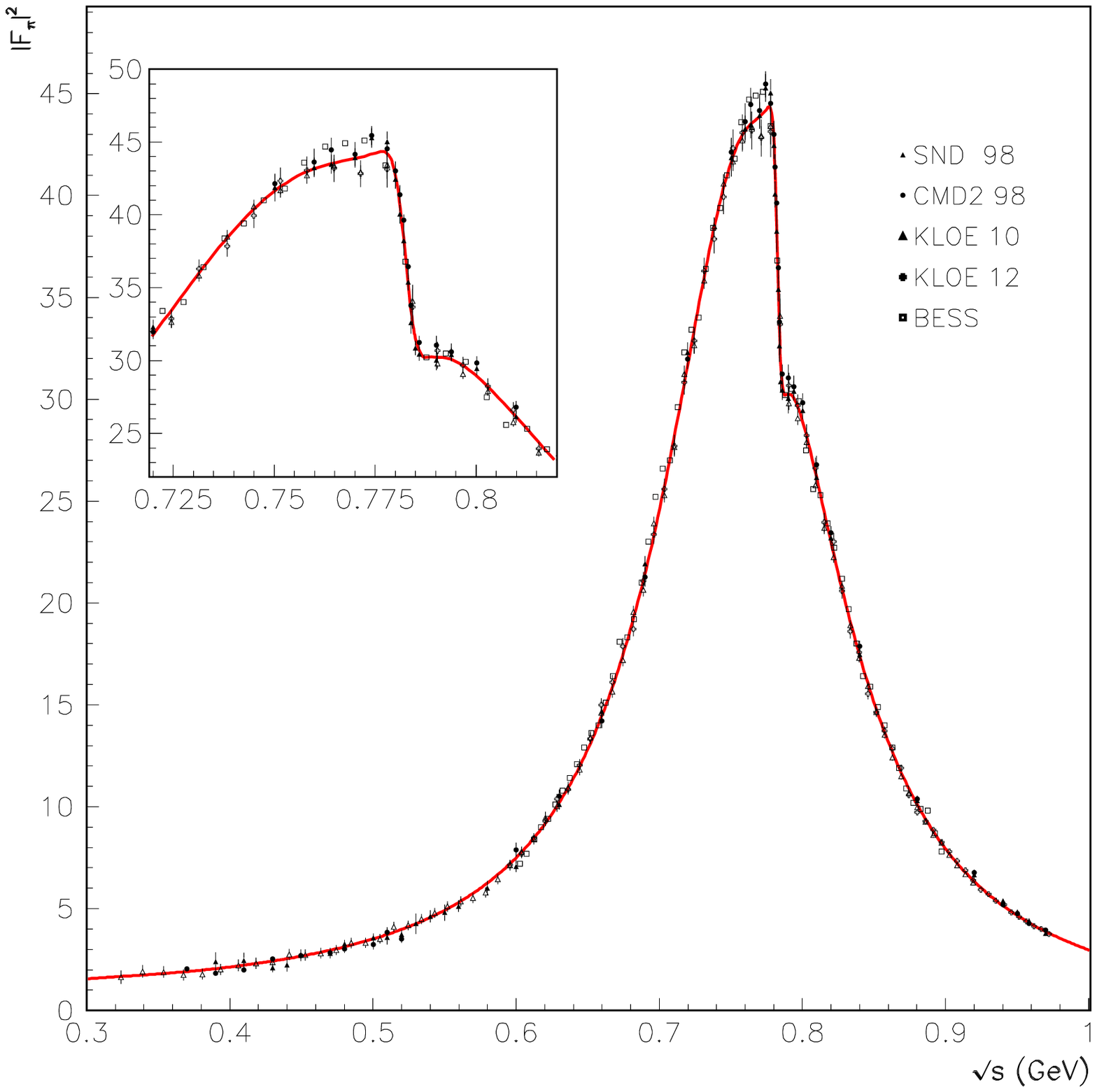}
\includegraphics[width=0.45\textwidth]{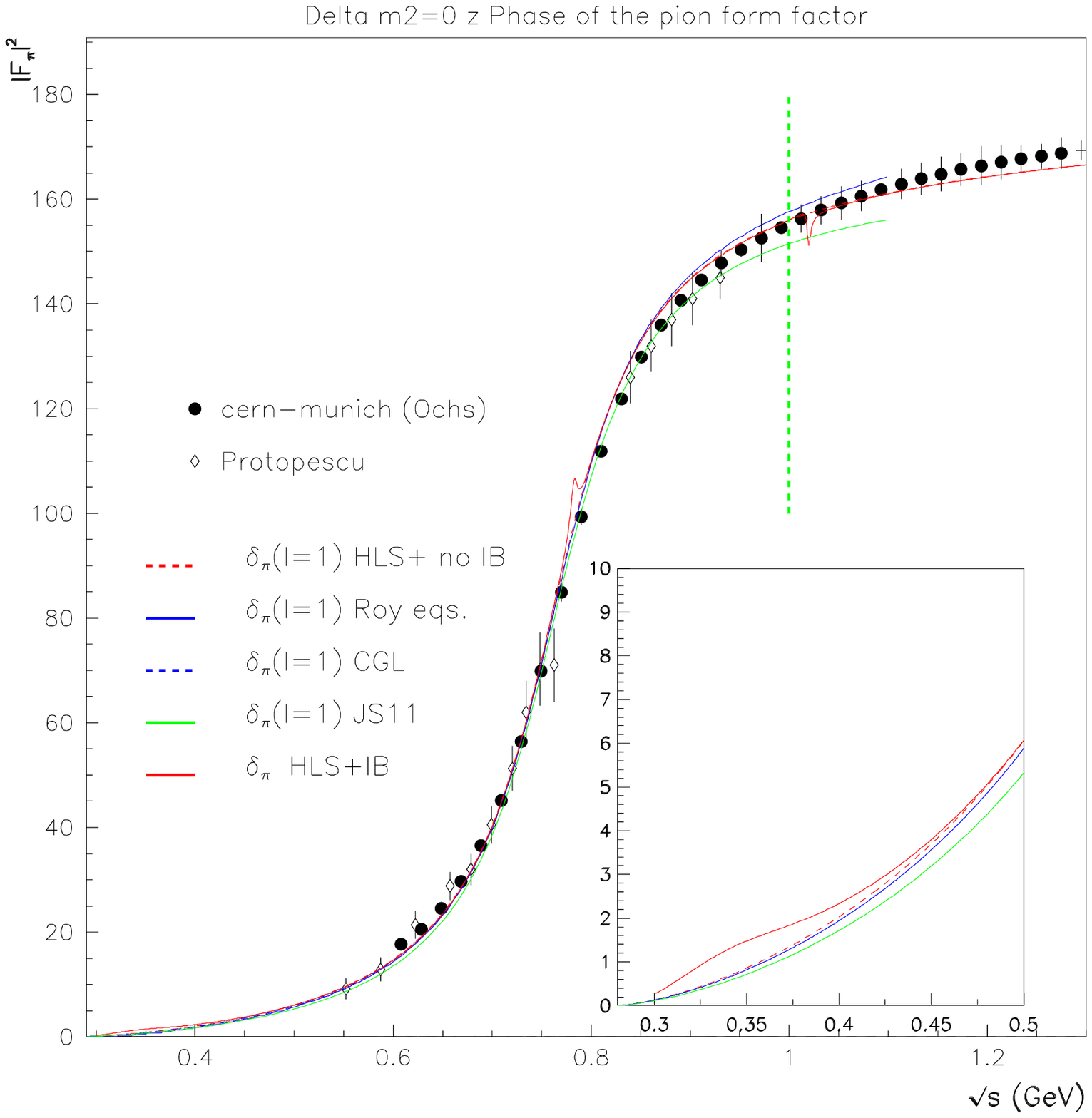}
\caption{The HLS model at work. The left panel shows the global
HLS model fit of the $\pi\pi$ channel together with
the data from Novosibirsk, Frascati and Beijing. The right panel shows
the $P$-wave $\pi^+ \pi^-$ phase-shift data and predictions from
\cite{pipiscatt, pipiscatt_1, pipiscatt_2, pipiscatt_3} and \cite{JS11} together with 
the broken HLS phase-shift. [Reprinted with permission from 
Ref.~\cite{Benayoun:2015gxa}. Copyright (2015) by the European Physical Journal C (EPJ C)]}
\label{fig:pipifit} 
\end{figure}

The HLS model, supplemented by isospin and
SU(3) breaking effects, has been tested to work surprisingly well up to 1.05
GeV, just including the $\phi$ meson~\cite{Benayoun:2012wc,Benayoun:2015gxa}. 
This is illustrated in the left panel of Fig.~\ref{fig:pipifit} for the case of 
the pion form factor $F_\pi(E)$.
 
Another important check is a comparison of the $\pi\pi$ rescattering as obtained in 
the model of Refs.~\cite{Benayoun:2012wc,Benayoun:2015gxa} with data and
with results once obtained by Colangelo and Leutwyler in their from
first principles approach~\cite{pipiscatt, pipiscatt_1, pipiscatt_2, pipiscatt_3}. 
One of the key ingredients
in this approach is the strong interaction phase shift $\delta^1_1(s)$
of $\pi\pi$ (re)scattering in the final state. We compare the phase of
$F_\pi(s)$ in our model with the one obtained by solving the Roy
equation with $\pi\pi$-scattering data as input.  We notice that the
agreement is surprisingly good up to about 1 GeV as shown in the right panel of
Fig.~\ref{fig:pipifit}.

The HLS model, which is an implementation of the vector meson dominance
(VMD) model in accord with the chiral structure of QCD, includes photonic
corrections treating hadrons as point-like particles. We have to address the
question of the range of validity of this model. 
A more precise understanding of the photon radiation by hadrons is
particularly important for the initial state radiation (ISR) radiative
return measurements of hadronic cross sections by KLOE, BaBar and
BES, which are based on sQED modelling, i.e., treating hadrons as point-like particles,
of the final state radiation (FSR).

\begin{figure}
\centering
\includegraphics[width=10cm]{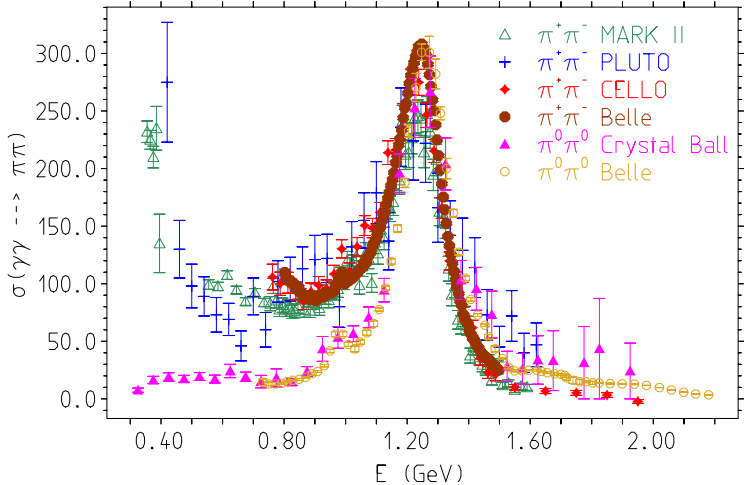}
\caption{How do photons couple to pions? This is obviously probed in 
reactions like $\gamma \gamma \to \pi^+\pi^-,\pi^0\pi^0$. One can infer from data 
that below about 1 GeV photons couple to pions as point-like objects
(i.e. only the charged pions are produced directly). At higher energies, the
photons see the quarks exclusively and form the prominent tensor
resonance $f_2(1270)$. The $\pi^0\pi^0$ cross section in this figure
is enhanced by the isospin symmetry factor 2, by which it is reduced
in reality.}
\label{fig:ggpipiall} 
\end{figure}

There is no doubt that sQED works at low energies if photons are
relatively soft. It has been in fact utilised to account for
the FSR corrections in processes involving the charged pions in the final state.
Direct experimental studies of the FSR spectrum at
intermediate energies, advocated
e.g. in~\cite{Gluza:2002ui,Pancheri:2006cp}, are not available yet
but, as far as studies exist, they seem to support
sQED~\cite{FBasymmetryKLOE, Ambrosino:2006hb}. The latter, however,
obviously has to break down in the hard photon regime. Here, di-pion
production in $\gamma \gamma$ fusion is able to shed more light on
that problem. Di-pion production cross sections are available from Crystal Ball, 
Mark II, JADE, PLUTO, CELLO and Belle~\cite{CBggpipi, Bienlein:1992em, JADEggpipi, Boyer:1990vu, PLUTOggpipi, CELLOggpipi, Belleggpipi, Uehara:2009cka, Uehara:2008pf, Belleggpi0pi0}. 
We see in Fig.~\ref{fig:ggpipiall} that the $\pi^+\pi^-$ cross section is large 
while the $\pi^0\pi^0$ one is tiny at threshold, which means that,
as expected, photons see the pions and they do not see the composite structure as
they are not hard enough. The $\pi^0\pi^0$ final state is then available via strong
rescattering only.

As energy of the $\pi\pi$ system increases, the strong tensor meson
resonance $f_2(1270)$ shows up in both the charged and the neutral
channels. Rates only differ by the isospin weight factor 2. Apparently
now photons directly probe the quarks. Figure~\ref{fig:ggpipiall} also
illustrates that utilising isospin relations to evaluate missing
contributions to $a_\mu^{\rm had,LO}$ from unseen channels may be
rather misleading, since we are dealing with hadron production
mediated by one photon exchange and electromagnetic interaction
obviously can violate isospin by close to 100\%.

In the present work, we present a sample of results for the cross sections of
several processes of low energetic $\epm$ annihilation into
final states containing pions accompanied by one or two photons, or a light
lepton pair. The results, which have been obtained with a new version
of a multipurpose Monte Carlo program {\tt carlomat}, labelled 3.1
\cite{carlomat3.1}, demonstrate new capabilities of the program.
They include a possibility of taking into account either the initial or final
state radiation separately, or both at a time, and a possibility of inclusion of
the electromagnetic charged pion form factor for processes with charged pion pairs.
We also discuss some problems related to the $U(1)$ electromagnetic gauge
invariance.

\section{Theoretical framework}

\begin{figure}[htb]
\centerline{
\epsfig{file=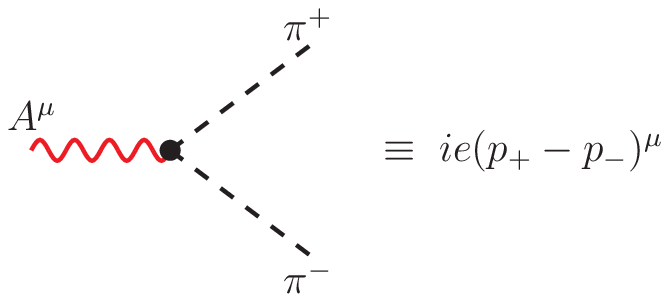,  width=40mm, height=20mm}
\hspace*{1cm}
\epsfig{file=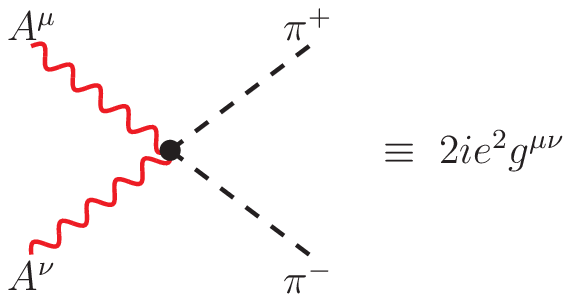,  width=40mm, height=20mm}}
\caption{\small Vertices of sQED}
\label{vertsQED}
\end{figure}

The production of charged pion pairs in the $\epm$ annihilation at low energies
can be effectively described in the framework of scalar quantum electrodynamics 
(sQED), with the $U(1)$ gauge invariant Lagrangian given by
\bea
\label{sqed}
\mathcal{L}_{\pi}^{\rm sQED}=
\partial_{\mu}\varphi\left(\partial^{\mu}\varphi\right)^*-m_{\pi}^2\varphi\varphi^*
-ie\left(\varphi^*\partial_{\mu}\varphi-\varphi\partial_{\mu}\varphi^*\right)
A^{\mu}
+e^2g_{\mu\nu}\varphi\varphi^*A^{\mu}A^{\nu},
\eea
where $\pi^{\pm}$ are represented by a complex scalar field $\varphi$ and the remaining
notation is obvious. The interaction vertices following from the Lagrangian (\ref{sqed}) 
are shown in Fig.~\ref{vertsQED}.

\begin{figure}[htb]
\centerline{
\includegraphics[width=0.5\textwidth]{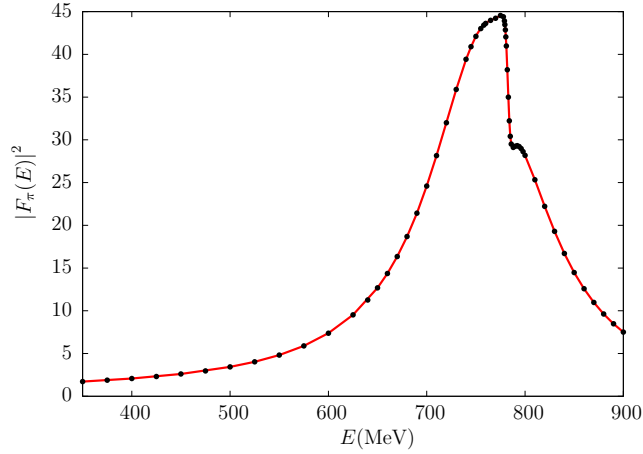}}
\caption{The EM charged pion form factor calculated by direct calls to
the subroutine for $F_{\pi}(s)$ ({\em points}) and with the MC program generated automatically
({\em solid line}).}
\label{pionff}
\end{figure}

The bound state nature of the charged pion can be taken into account
by introducing in a proper way the charged pion form factor
$F_{\pi}(q^2)$ in the Feynman rules of Fig.~\ref{vertsQED}. In the
time-like region, $q^2=s > 4m_{\pi}^2$, the form factor is given by
\bea
\label{piff}
\left|F_{\pi}(s)\right|^2=\frac{\sigma^{(0)}(e^+e^-\to\gamma^*\to\pi^+\pi^-)}
{\frac{\pi\alpha^2}{3s}\beta_{\pi}^3}, \qquad {\rm with}\quad 
\beta_{\pi}=\left(1-\frac{4m_{\pi}^2}{s}\right)^{\frac{1}{2}}.
\eea
Actually, the HLS model predicts the pion form factor, but one has to include
one-loop self-energy corrections in order to account for the dynamically
generated width of the $\rho$ meson and its energy dependence. 
A separate subroutine for the form factor $F_{\pi}(s)$, based on a fit to
data from Novosibirsk \cite{NSK1, NSK2, NSK3, NSK4, NSK5}, Frascati (KLOE) 
\cite{KLOE1, KLOE1_1, KLOE2, KLOE3}, SLAC (BaBar) \cite{BaBar, BaBar1} and Beijing (BESIII)
\cite{BESIII}, has been written by one of us (FJ). 
It has been then implemented in {\tt carlomat} by the substitution:
\bea
\label{subst3}
e\to eF_{\pi}(q^2)
\eea
in the triple coupling of Fig.~\ref{vertsQED} and appropriate modification of the
quartic coupling of Fig.~\ref{vertsQED}, which will be discussed later. Correctness 
of the implementation is cross checked by comparing the values of the form factor 
calculated by a direct call to the subroutine for $F_{\pi}(s)$ against corresponding values 
calculated according to Eq.~(\ref{piff}), 
i.e., with the MC program for $\sigma^{(0)}(e^+e^-\to\gamma^*\to\pi^+\pi^-)$ generated 
automatically with {\tt carlomat\_3.1}. The results of the cross check are plotted 
in Fig.~\ref{pionff}.

\begin{figure}[htb]
\centerline{
\epsfig{file=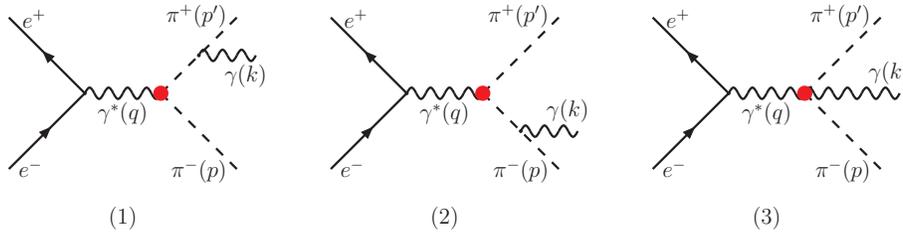,  width=120mm, height=30mm}}
\caption{\small FSR Feynman diagrams of process (\ref{ppa}) in the LO. The relevant four
momenta are indicated in parentheses and blobs represent the charged pion form factor.}
\label{ee2ppa_fsr}
\end{figure}

\begin{figure}[htb]
\centerline{
\epsfig{file=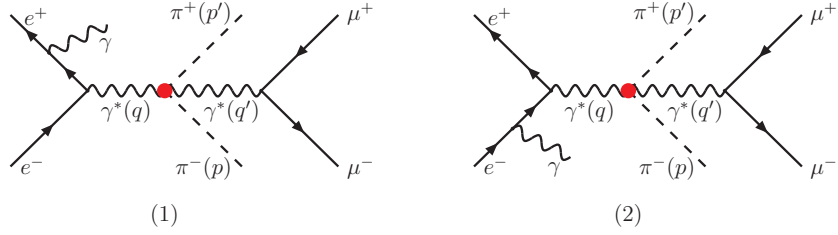,  width=110mm, height=30mm}}
\caption{\small ISR Feynman diagrams of process (\ref{ppmma}) in which the choice
of four momentum transfer in the charged pion form factor, represented
by the {\em red blob}, may be ambiguous.}
\label{ee2ppmma_isr}
\end{figure}

As the fit formula for $F_{\pi}(s)$ includes contributions 
of the virtual photon mixing with $\rho^0$, $\omega$, $\phi$, $\rho(1450)$ and 
$\rho(1700)$ vector mesons, each subsequently decaying into the 
$\pi^+\pi^-$-pair, substitution (\ref{subst3}) is not justified if a real photon is radiated
off the final state pion. In this case, it seems better to keep the triple coupling 
of Fig.~\ref{vertsQED} unchanged. However, then a question arises, how the form factor 
should be taken into account in the quartic coupling of Fig.~\ref{vertsQED} in
order not to violate the $U(1)$ gauge invariance.
To address the question let us consider the FSR in the process
\bea
\label{ppa}
e^+e^-\to\pi^+\pi^-\gamma
\eea
the LO Feynman diagrams of which are shown in Fig.~\ref{ee2ppa_fsr}, where the particle
four momenta relevant for the Feynman rules of Fig.~\ref{vertsQED} have been indicated 
in parentheses.
Using the substitutions: (\ref{subst3}) in the triple and $e^2\to e^2G_{\pi}(q^2)$ in 
the quartic vertex of Fig.~\ref{vertsQED}, we
get the following expressions for the amplitudes of the LO Feynman diagrams of 
Fig.~\ref{ee2ppa_fsr}:
\bea
\label{M1}
M_1&=&j_{\mu}eF_{\pi}(q^2)(p'+k-p)^{\mu}\;e\;\frac{\left(-p'-(p'+k)\right)\cdot
\varepsilon(k)}{(p'+k)^2-m_{\pi}^2},\\
\label{M2}
M_2&=&j_{\mu}eF_{\pi}(q^2)(p'-p-k)^{\mu}\;e\;\frac{\left((p+k)+p\right)\cdot
\varepsilon(k)}{(p+k)^2-m_{\pi}^2},\\
\label{M3}
M_3&=&j_{\mu}2e^2G_{\pi}(q^2)g^{\mu\nu}\varepsilon_{\nu}(k),
\eea
where $j$ stands for the initial state current contracted with the photon
propagator and we have suppressed polarisation indices both in $j$ and in the photon
polarisation four vector $\varepsilon(k)$. Now, in order to test the $U(1)$ gauge 
invariance, let us substitute $\varepsilon(k)\to k$ in the full FSR amplitude
\bea
\label{MFSR}
\left.M_{\rm FSR}\right|_{\varepsilon(k)\to k}&=&
\left.\left(M_1+M_2+M_3\right)\right|_{\varepsilon(k)\to k}\nn\\
&=& j_{\mu}e^2\left[(p'+k-p)^{\mu}F_{\pi}(q^2)\frac{-2p'\cdot k}{2p'\cdot k}
+(p'-p-k)^{\mu}F_{\pi}(q^2)\frac{2p\cdot k}{2p\cdot k} + 2k^{\mu}G_{\pi}(q^2)\right]\nn\\
&=&j_{\mu}e^2\left[-2k^{\mu}F_{\pi}(q^2) + 2k^{\mu}G_{\pi}(q^2)\right].
\eea
It is obvious that the right hand side of Eq.~(\ref{MFSR}) vanishes only if
$G_{\pi}(q^2)\equiv F_{\pi}(q^2)$. Thus, the EM charged pion form factor should be included 
in the quartic vertex of Fig.~\ref{vertsQED} by the following substitutions:
\bea
\label{subst4}
e^2\to e^2F_{\pi}(q^2), \qquad e^2\to e^2\left|F_{\pi}(q^2)\right|^2,
\eea
if one, or none, respectively, of the photon lines is on mass shell. Both possibilities
of modification of $e^2$
are included in the program at the stage of code generation and can be controlled in 
the MC computation part of the program. However, the situation, where both photons 
in the vertex are off-shell, may lead to some ambiguity of the choice of
the momentum transfer in the form factor. This is illustrated in Fig.~\ref{ee2ppmma_isr}, 
where two initial state radiation (ISR) Feynman diagrams of the process
\bea
\label{ppmma}
e^+e^-&\to&\pi^+\pi^-\mu^+\mu^-\gamma
\eea
are shown. It is not at all clear, which four momentum, $q$ or $q'$, should be
used in the charged pion form factor that is to be substituted in the quartic sQED vertex
indicated by the blob. The non-trivial pion form factor is
due to the VMD dressing of the photon via $\rho$--$\gamma$ mixing, and in
fact the HLS model exhibits a $\rho^0\rho^0 \pi^+\pi^-$ coupling as
well. Thus both off-shell photons are dressed with a form
factor. As $q^2$ is tuned to scan the $\rho$ resonance the
corresponding form factor is crucial as it exhibits large deviations
from unity in the resonance region, while $q^{'2}$ is closer to the
photon mass shell such that the form factor is much less important
 as we expect to see only the low energy tail of the resonance. A
full HLS model calculation is expected to remove any ambiguity here.
However, in {\tt carlomat\_3.1}, the choice of four momentum in the form factor 
is made automatically and it need not be consistent between
same vertices appearing in different Feynman diagrams. Such 
inconsistencies may lead to violation of the $U(1)$ gauge invariance. 
The inconsistency may also occur if we want to treat the ISR in process (\ref{ppmma})
in an inclusive way according to Eq.~(1) of Ref.~\cite{bkm}, where
the cross section of non-radiative process 
\bea
\label{ppmm}
e^+e^-&\to&\pi^+\pi^-\mu^+\mu^-
\eea
should be folded with the corresponding radiation function describing the ISR. Again, 
it may happen 
that the pion form factor in the Feynman diagram of process (\ref{ppmm}), which can be 
obtained either from diagram (1) or (2) of Fig.~\ref{ee2ppmma_isr}
by cancelling the external photon line, will be parameterised in terms of different
four momenta relative to the corresponding form factor in process (\ref{ppmma}). Needless
to say, such ambiguities in the choice of four momenta in the form factor may lead to
substantial discrepancies between the corresponding cross sections. 
Therefore, it is better to use a fixed coupling in such cases.

\begin{figure}[htb]
\centerline{
\epsfig{file=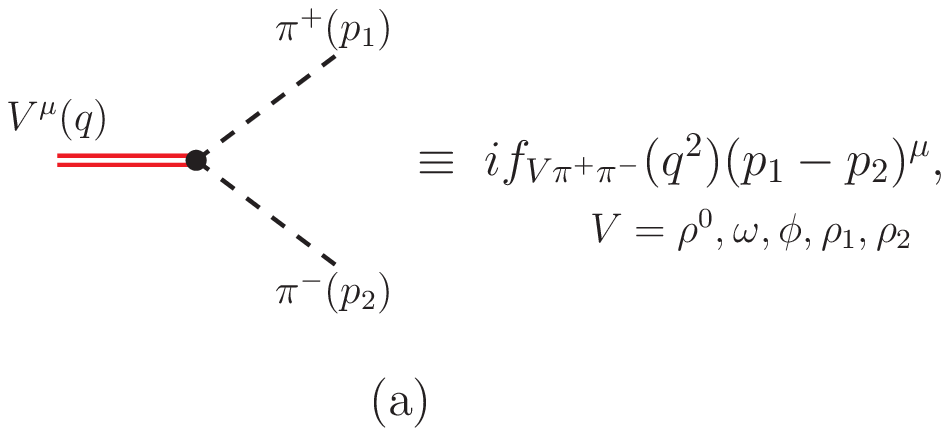,  width=60mm, height=30mm}
\hspace*{1cm}
\epsfig{file=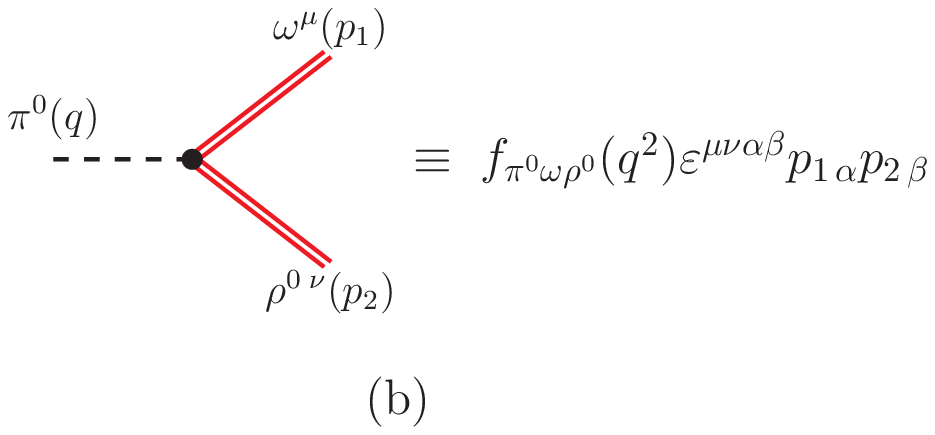,  width=60mm, height=30mm}}
\vspace*{4mm}
\centerline{
\epsfig{file=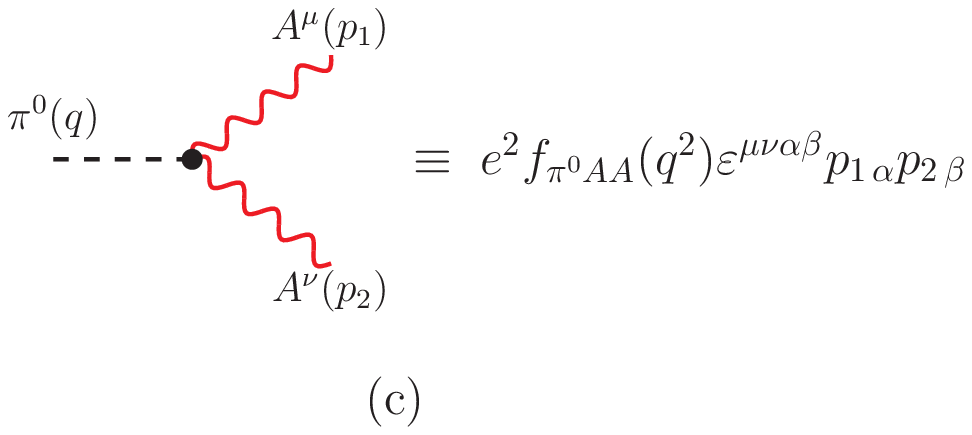,  width=60mm, height=30mm}
\hspace*{1cm}
\epsfig{file=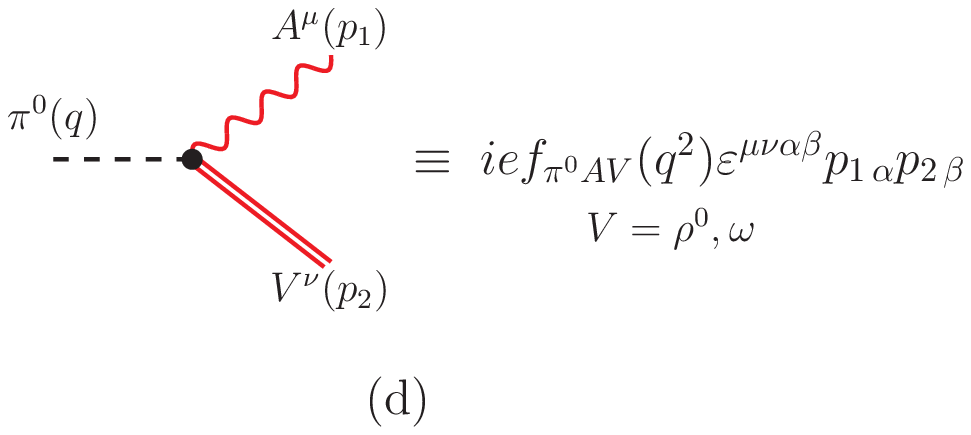,  width=60mm, height=30mm}}
\caption{\small Triple vertices of the HLS relevant for the present work.}
\label{vert3}
\end{figure}

\begin{figure}[htb]
\centerline{
\epsfig{file=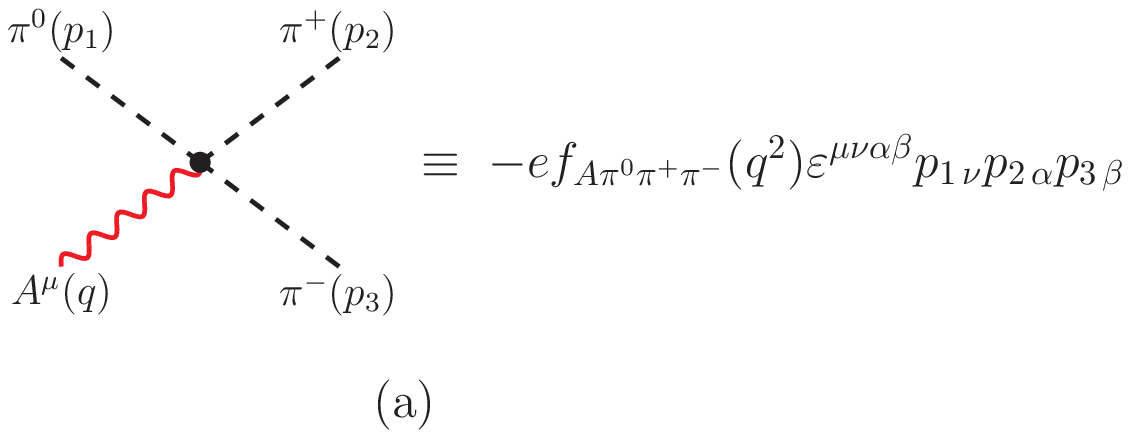,  width=70mm, height=30mm}
\hspace*{1cm}
\epsfig{file=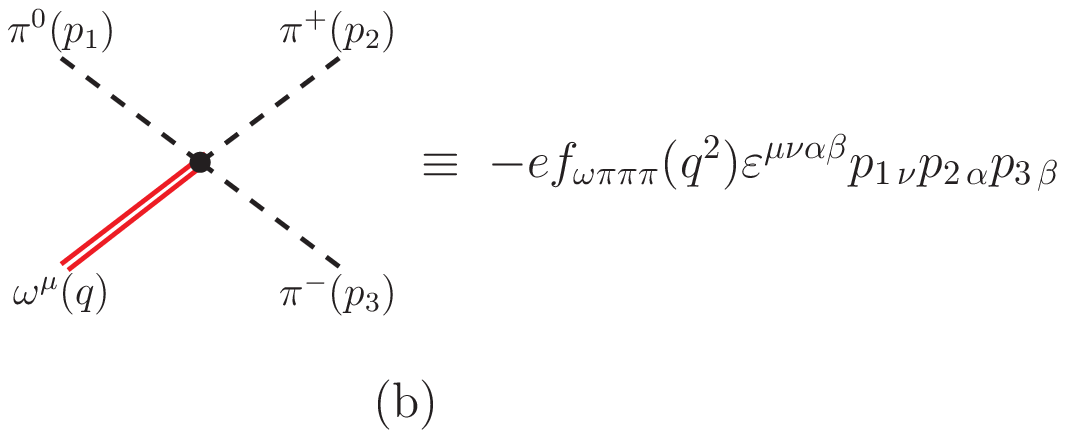,  width=70mm, height=30mm}}
\vspace*{4mm}
\centerline{
\epsfig{file=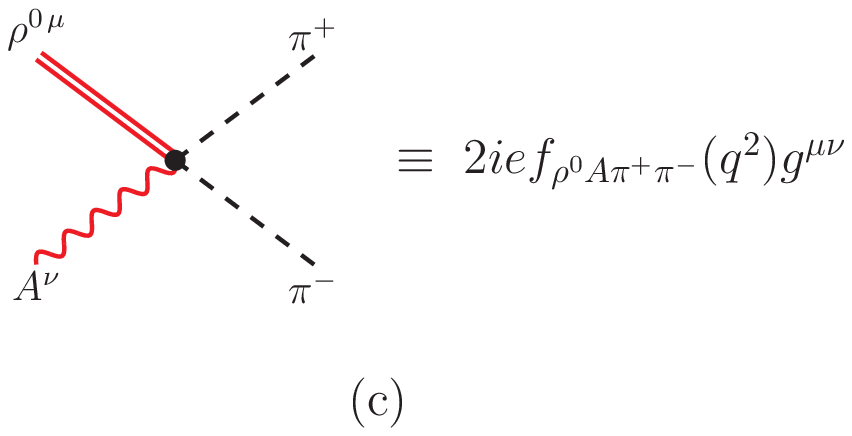,  width=60mm, height=30mm}}
\caption{\small Quartic vertices of the HLS model taken into account in the present work.}
\label{vert4}
\end{figure}

In order to describe the processes of pion production in $\epm$ annihilation 
at low energies, we will use, in addition to the SM and sQED vertices of 
Fig.~\ref{vertsQED},
\begin{figure}[htb]
\centerline{
\epsfig{file=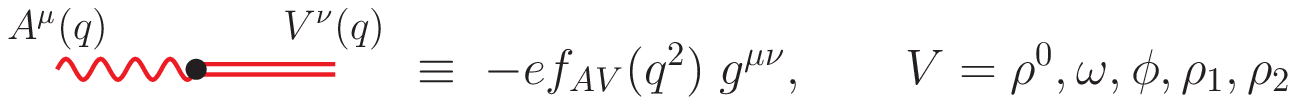,  width=100mm, height=7mm}}
\caption{\small The $\gamma$---$V$-mixing terms considered in the present work;
$\rho_1$ and $\rho_2$ stand for $\rho(1450)$ and $\rho(1700)$, respectively.}
\label{mixing}
\end{figure}
if appropriate with substitutions (\ref{subst3}) and (\ref{subst4}), the photon--vector 
meson mixing of Fig.~\ref{mixing} and the triple and quartic vertices of the HLS model 
depicted, respectively, in Figs. \ref{vert3} and \ref{vert4}. 
In Figs~\ref{mixing}
and \ref{vert3}a, $\rho_1$ and $\rho_2$ stand for $\rho(1450)$ and $\rho(1700)$, 
respectively.
If we include the EM charged pion form factor, then we neglect 
the contributions of the $\gamma$--$V$ mixing
and the subsequent decay of the corresponding vector meson into $\pi^+\pi^-$-pairs, 
represented by the interaction vertex of Fig.~\ref{vert3}a, as well as the
contribution of the quartic vertex of Fig.~\ref{vert4}c, as they are already
included in the form factor.

\section{Results}

\begin{figure}[htb]
\centerline{
\epsfig{file=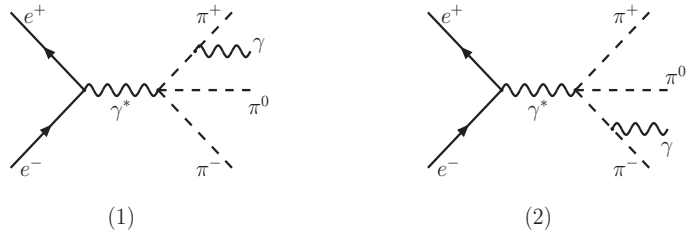,  width=90mm, height=30mm}}
\caption{\small The Feynman diagrams of process (\ref{pppa}) that cause the $U(1)$ 
gauge invariance violation.}
\label{diags_pppa}
\end{figure}

In this section, we show a sample of results for the cross sections of several
processes of $\epm\to$~hadrons, with the final state containing pions accompanied
by one or two photons, or a light lepton pair, at $\sqrt{s}=0.8$~GeV,
$\sqrt{s}=1.0$~GeV and $\sqrt{s}=1.5$~GeV. The necessary MC programs have been
generated taking into account the Feynman rules described in Sect.~2
and computed with the same physical input
parameters as those specified in module {\tt inprfms} of {\tt carlomat\_3.1}
which is available on the web page \cite{carlomat3.1}. The following cuts on
the photon energy $E_{\gamma}$ and photon angle with respect to the beam
$\theta_{\gamma {\rm b}}$ are imposed:
\bea
\label{cuts}
E_{\gamma}>    0.01\; {\rm GeV},
\qquad     5^{\circ} < \theta_{\gamma {\rm b}} <  175^{\circ}.
\eea
In addition to processes (\ref{ppa}) and (\ref{ppmma}), we consider 
the following radiative processes with one photon:
\bea
\label{pppa}
e^+e^-&\to&\pi^+\pi^-\pi^0\gamma,\\
\label{ppppa}
e^+e^-&\to&\pi^+\pi^-\pi^+\pi^-\gamma
\eea
and two photons:
\bea
\label{ppaa}
e^+e^-&\to&\pi^+\pi^-\gamma\gamma,\\
\label{ppmmaa}
e^+e^-&\to&\pi^+\pi^-\mu^+\mu^-\gamma\gamma,\\
\label{ppppaa}
e^+e^-&\to&\pi^+\pi^-\pi^+\pi^-\gamma\gamma.
\eea
The cross sections of processes (\ref{ppa}), (\ref{ppmma}) and
(\ref{pppa})--(\ref{ppppaa}) at $\sqrt{s}=1$~GeV, computed with cuts (\ref{cuts}),
are listed in Table~1, where each row includes the ISR and 
full LO total cross sections together with results of the $U(1)$ gauge invariance 
tests, i.e. the corresponding cross sections computed with the photon polarisation
four vector $\varepsilon(k)$ replaced with its four momentum $k$. 
The replacement is made just for one photon for processes
(\ref{ppaa})--(\ref{ppmmaa}) with two photons. For testing purposes, we also give
corresponding numbers of the Feynman diagrams which {\tt carlomat\_3.1} generated
for each of the considered processes within the model with the EM charged pion form
factor, except for process (\ref{pppa}) where we give the number of diagrams
generated within the HLS model with fixed couplings, as specified in Sect.~2.
The gauge invariant test is satisfied perfectly well for all the ISR cross sections 
presented and full LO cross sections of processes (\ref{ppa}), (\ref{ppmma}),
(\ref{ppppa}) and (\ref{ppaa}), for which we observe a drop of about 32 orders
of magnitude, in accordance to what can be expected with the double precision
Fortran arithmetic. A less satisfactory drop of the full LO cross section of
processes (\ref{ppmmaa}) and (\ref{ppppaa}) is due to the ambiguity of the
momentum transfer choice in the pion form factor $F_{\pi}(q^2)$ in the Feynman
diagrams containing the quartic vertex of sQED with both photon lines 
being virtual, as discussed in Sect.~2.
However, the four momentum transfer choice ambiguity cannot explain a much less 
satisfactory drop in the full LO cross section of process (\ref{pppa}), presented
in the second row of Table~1, where the problem 
is caused among others by the two Feynman diagrams shown in Fig.~\ref{diags_pppa}.
Apparently the Feynman rules listed in Figs.~\ref{mixing}, \ref{vert3} and
\ref{vert4} are incomplete and there should be an extra Feynman diagram with
the external photon attached to the quartic $\pi^+\pi^-\pi^0\gamma$ vertex,
i.e. a penta vertex, which would have cured the problem if the penta coupling had
been chosen properly. Unfortunately, an inclusion of that kind of coupling is not possible
in the automatic code generation, as the Feynman diagram topology generator of {\tt carlomat}
\cite{carlomat} includes only triple and quartic interaction vertices. However, it should
be noted that the $U(1)$ gauge invariance violation concerns only the FSR part which is
substantially smaller than the ISR part of the cross section. Thus, we expect that the
prediction for the cross section of (\ref{pppa}) can be still considered as being
quite reliable.

It should be stressed here that {\tt carlomat\_3.1} offers a possibility to compute
cross sections of other processes, e.g., with higher pion multiplicities, with neutral
pions, or with charged or neutral kaons, for which a similar kind of analyses could
easily be repeated.

\begin{table}
\begin{center}
\begin{tabular}{lrllcrll}
\hline
{\tt Process} & \multicolumn{3}{c}{\tt ISR} && \multicolumn{3}{c}{\tt Full LO}\\[1mm]
               \cline{2-4}                  \cline{6-8} 
$\epm\to$     & \# diags & $\sigma $ & $\sigma|_{\varepsilon(k)=k}$ &
              & \# diags & $\sigma $ & $\sigma|_{\varepsilon(k)=k}$ \\[1mm]
\hline
$\pi^+\pi^-\gamma$                 
& 2 $\;$& 2.041(4)e+4 & 1.04(1)e--28 && 5 $\;$& 2.249(4)e+4 & 2.73(2)e--28 \\
$\pi^+\pi^-\pi^0\gamma$            
& 32 $\;$& 409(1) & 2.21(3)e--30     && 156 $\;$& 481.5(6) & 3.011(1)e--2 \\
$\pi^+\pi^-\mu^+\mu^-\gamma$       
& 26 $\;$& 4.344(9)e--2 & 4.62(5)e--34 && 107 $\;$& 6.449(8)e--2 & 6.42(5)e--34\\
$\pi^+\pi^-\pi^+\pi^-\gamma$       
& 36 $\;$& 2.029(5)e--3 & 2.14(3)e--35 && 200 $\;$& 3.320(5)e--3 & 3.03(2)e--35 \\
$\pi^+\pi^-\gamma\gamma$           
& 6 $\;$& 1.445(14)e+3 & 1.22(4)e--29 && 44 $\;$& 2.131(8)e+3 & 2.08(3)e--29 \\
$\pi^+\pi^-\mu^+\mu^-\gamma\gamma$ 
& 90 $\;$& 1.127(7)e--3 & 1.16(4)e--35 && 1272 $\;$& 2.535(8)e--3 & 9.56(1)e--19\\
$\pi^+\pi^-\pi^+\pi^-\gamma\gamma$ 
& 120 $\;$& 4.68(3)e--5 & 4.6(1)e--37 && 2772 $\;$& 1.303(4)e--4 & 4.969(4)e--15\\
\hline
\end{tabular}
\caption{\small The cross sections in pb at $\sqrt{s}=1$~GeV and the corresponding
$U(1)$ gauge invariance tests. The numbers in parentheses show the MC uncertainty
of the last decimals.}
\end{center}
\label{tabcs}
\end{table}

\begin{figure}[htb]
\centerline{
\epsfig{file=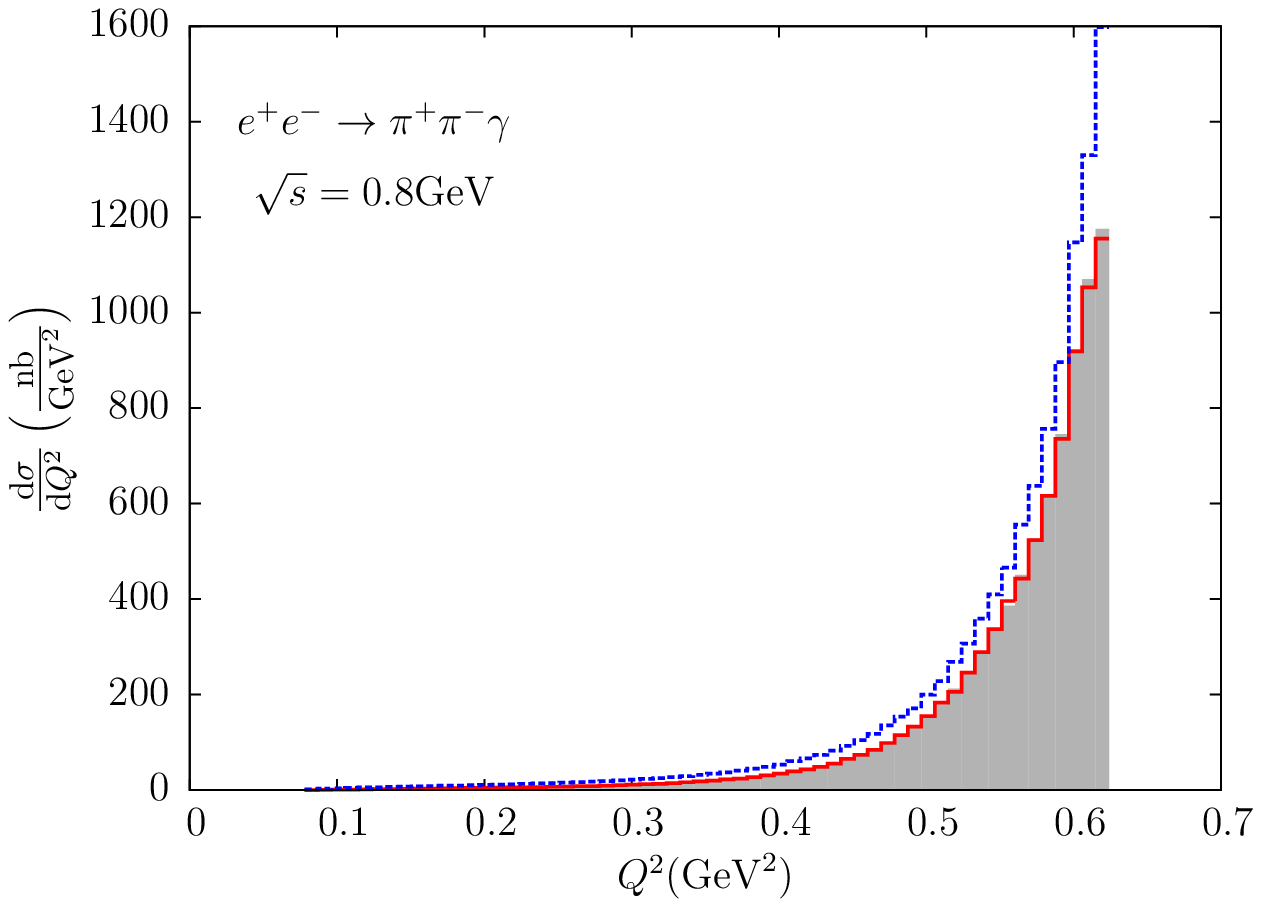,  width=55mm, height=42mm}
\epsfig{file=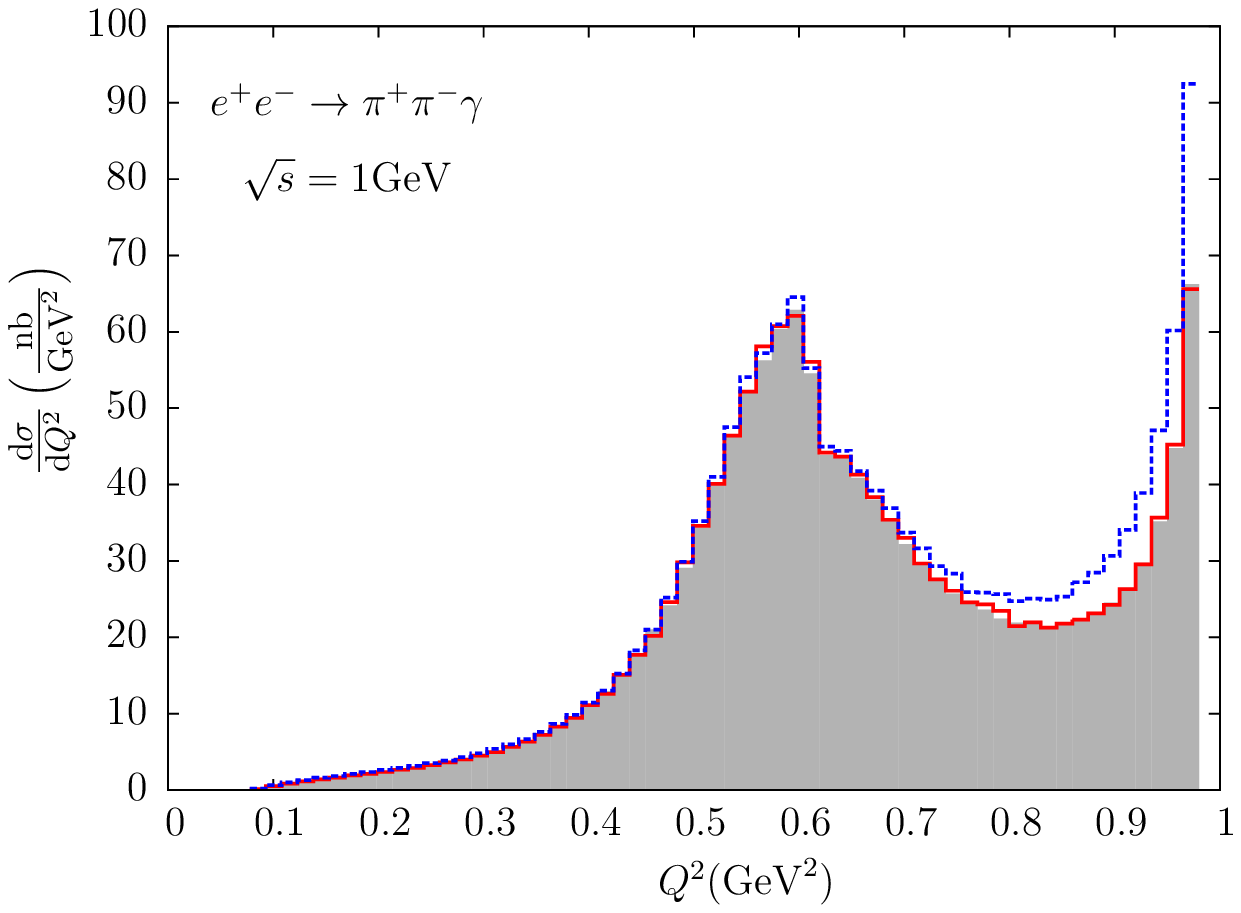,  width=55mm, height=42mm}
\epsfig{file=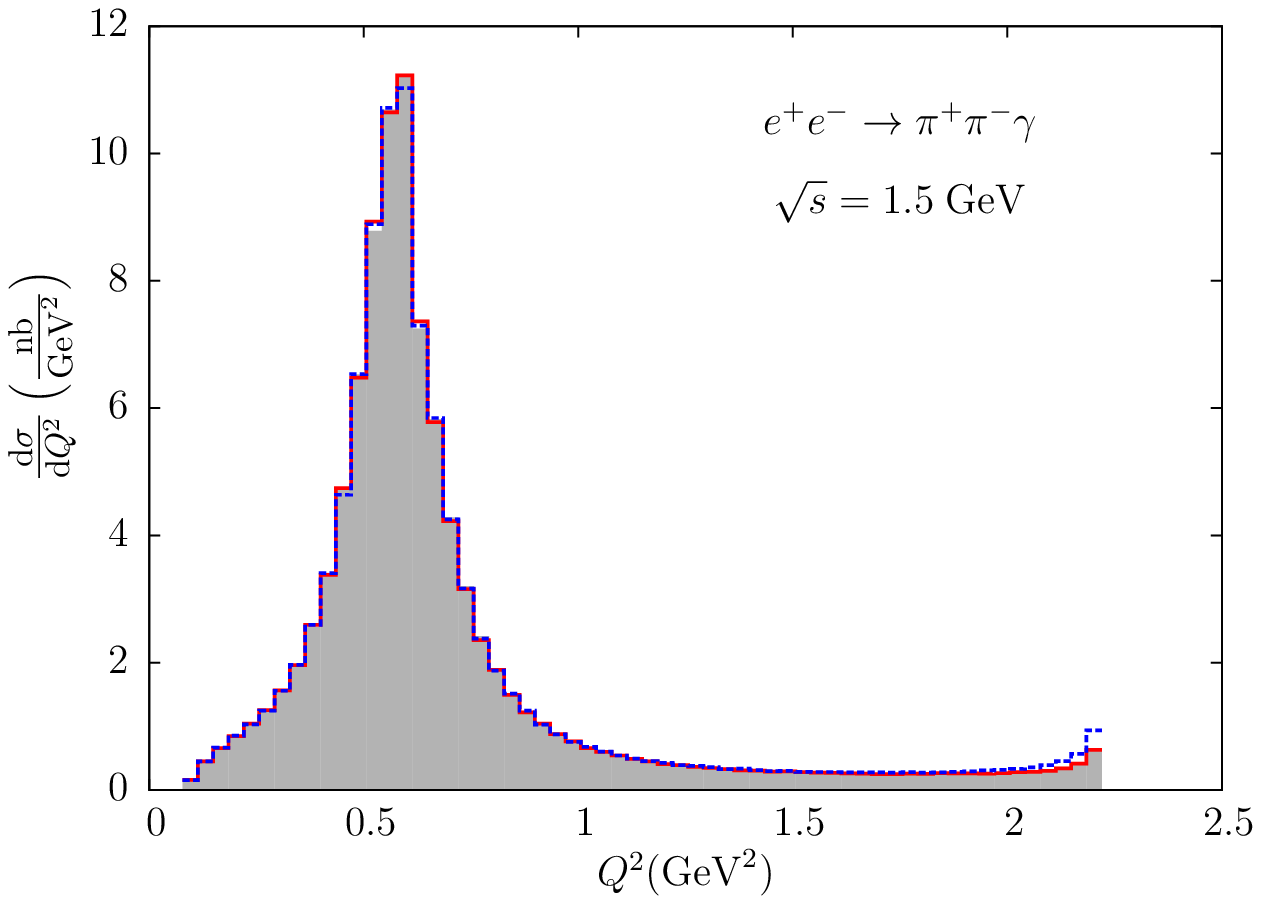,  width=55mm, height=42mm}}
\caption{\small The differential cross sections of process (\ref{ppa}) as functions
of invariant mass of the $\pi^+\pi^-$-pair. The solid line and the shaded histogram
represent the ISR cross section, obtained with the MC program and with the
analytic formula of Ref.~\cite{bkm}, respectively, and the dashed line represents the
full LO result. The difference between solid and dashed lines represents the
FSR correction.}
\label{pp_dQ2}
\end{figure}

The differential cross sections of process (\ref{ppa}) at $\sqrt{s}=0.8$~GeV, 
$\sqrt{s}=1$~GeV and $\sqrt{s}=1.5$~GeV are plotted in Fig.~\ref{pp_dQ2}
as functions of the invariant mass of the $\pi^+\pi^-$-pair. In all three panels,
the solid lines show the ISR cross section calculated with the MC program and 
the grey shaded histograms show the same cross section calculated with the
corresponding analytic formula for ${\rm d}\sigma_{\rm ISR}/{\rm d}Q^2$ given
by Eq.~(1) of ref.~\cite{bkm}. As can be seen in all the three panels, the two
differential cross sections agree perfectly well and small differences in a few
bins of $Q^2$ are most probably due to statistical fluctuations, as the
corresponding total cross sections agree within one standard 
deviation of the MC integration. 
The dashed lines show the full LO differential cross sections. Thus, the difference
between the dashed and solid lines illustrates the FSR effect.

\begin{figure}[htb]
\centerline{
\epsfig{file=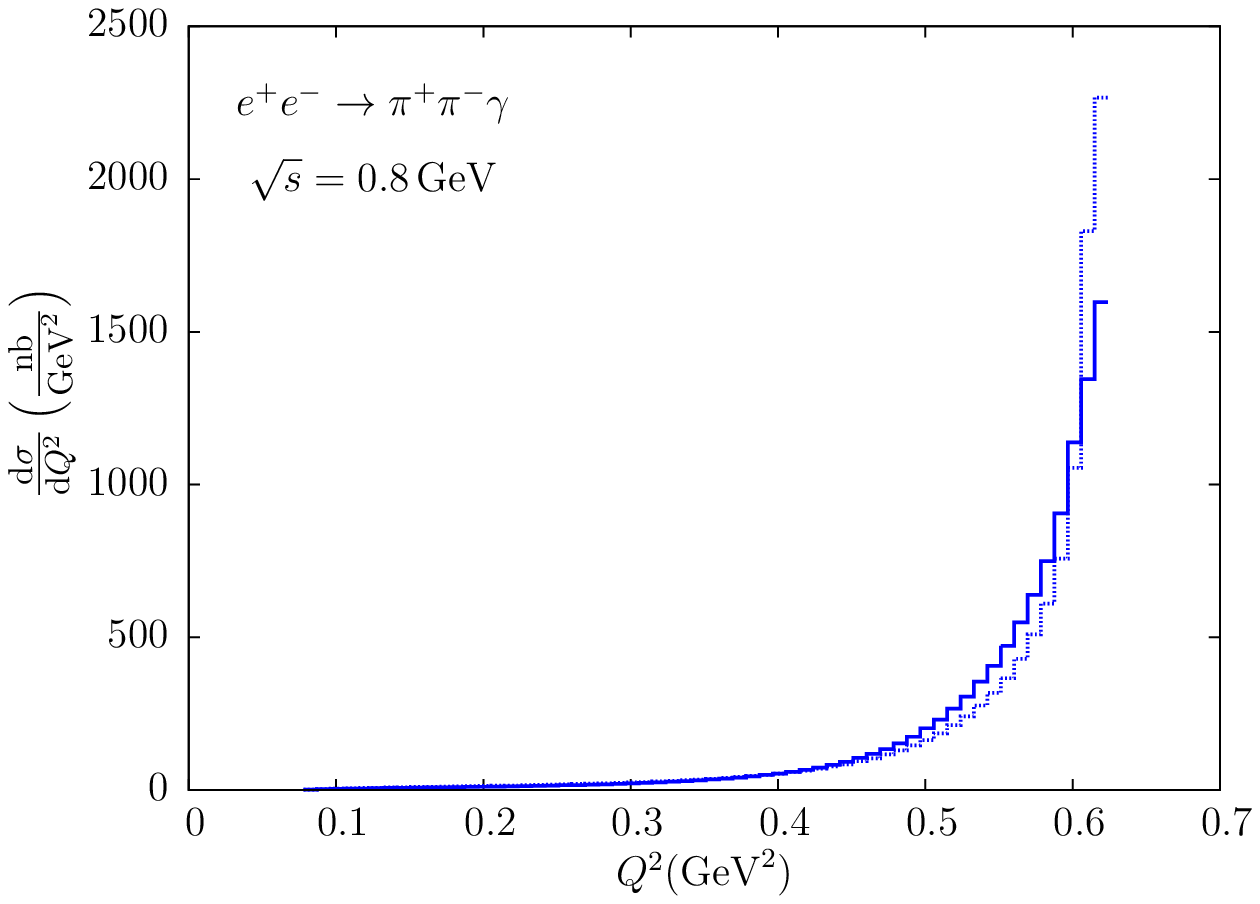,  width=55mm, height=42mm}
\epsfig{file=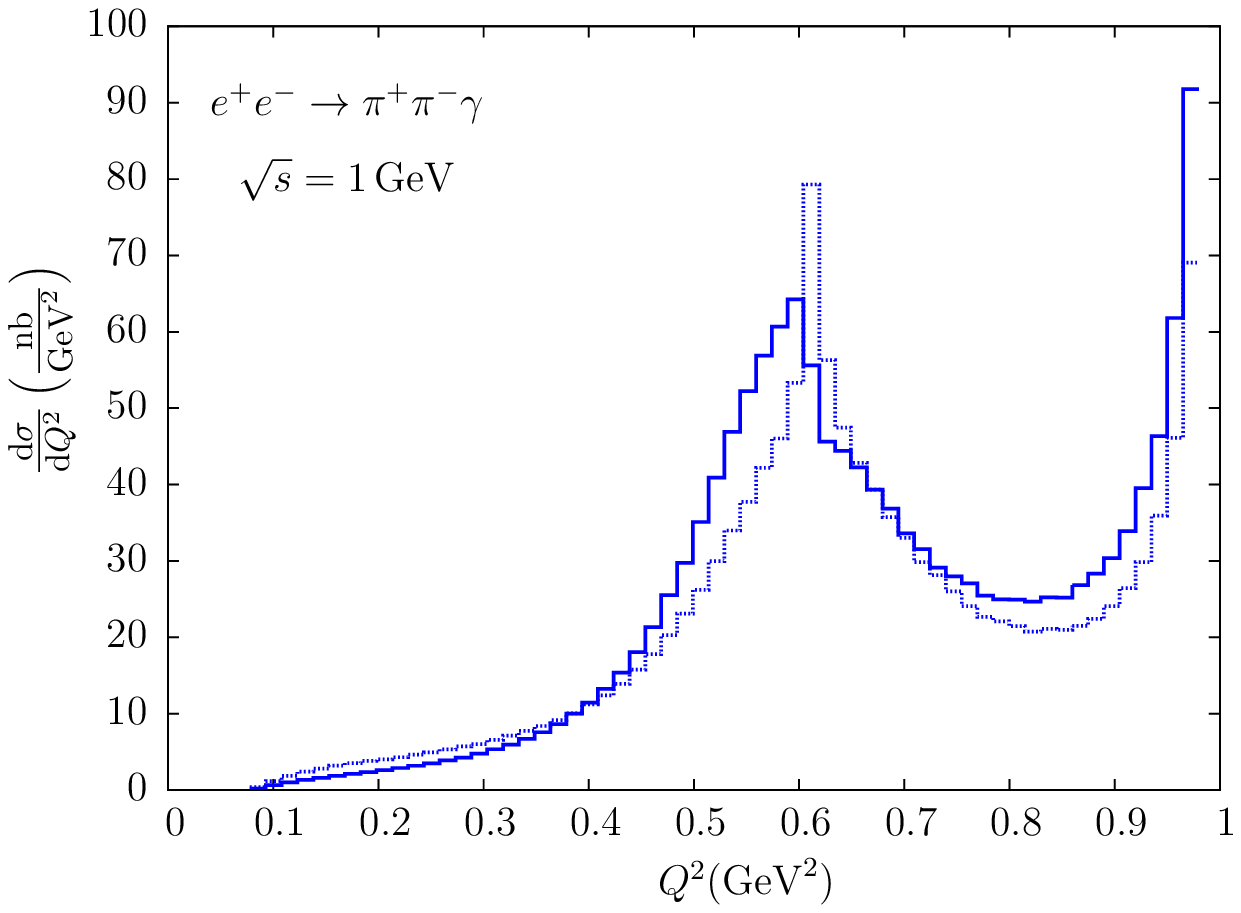,  width=55mm, height=42mm}
\epsfig{file=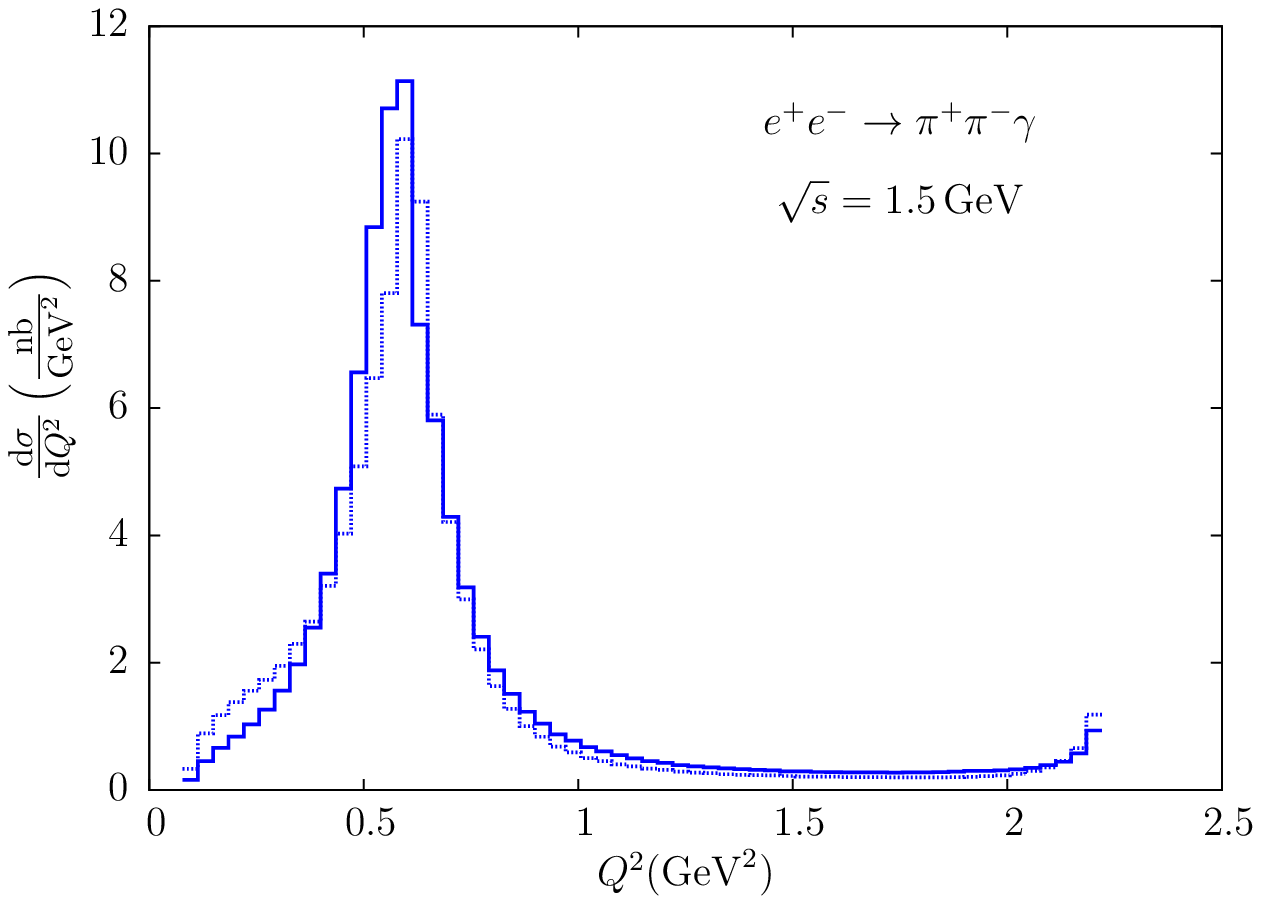,  width=55mm, height=42mm}}
\caption{\small The differential cross sections of (\ref{ppa}) as functions
  of invariant mass of the $\pi^+\pi^-$-pair computed in a model with the charged
  pion form factor ({\em solid lines}) and in the HLS model with fixed couplings ({\em dotted
  lines}).}
\label{ppa_ff_vs_diags}
\end{figure}

In order to illustrate the effect of the charged pion form factor,
we compare in Fig.~\ref{ppa_ff_vs_diags} the full LO differential cross sections
of process (\ref{ppa}), as plotted with the solid lines in Fig.~\ref{pp_dQ2},
with the corresponding LO cross sections calculated in the HLS model
defined by the Feynman rules of Figs. \ref{mixing}, \ref{vert3} and \ref{vert4}
with fixed couplings. The fixed coupling cross sections receive contribution from
26 Feynman diagrams. They are plotted with dotted lines.

\begin{figure}[htb]
\centerline{
\epsfig{file=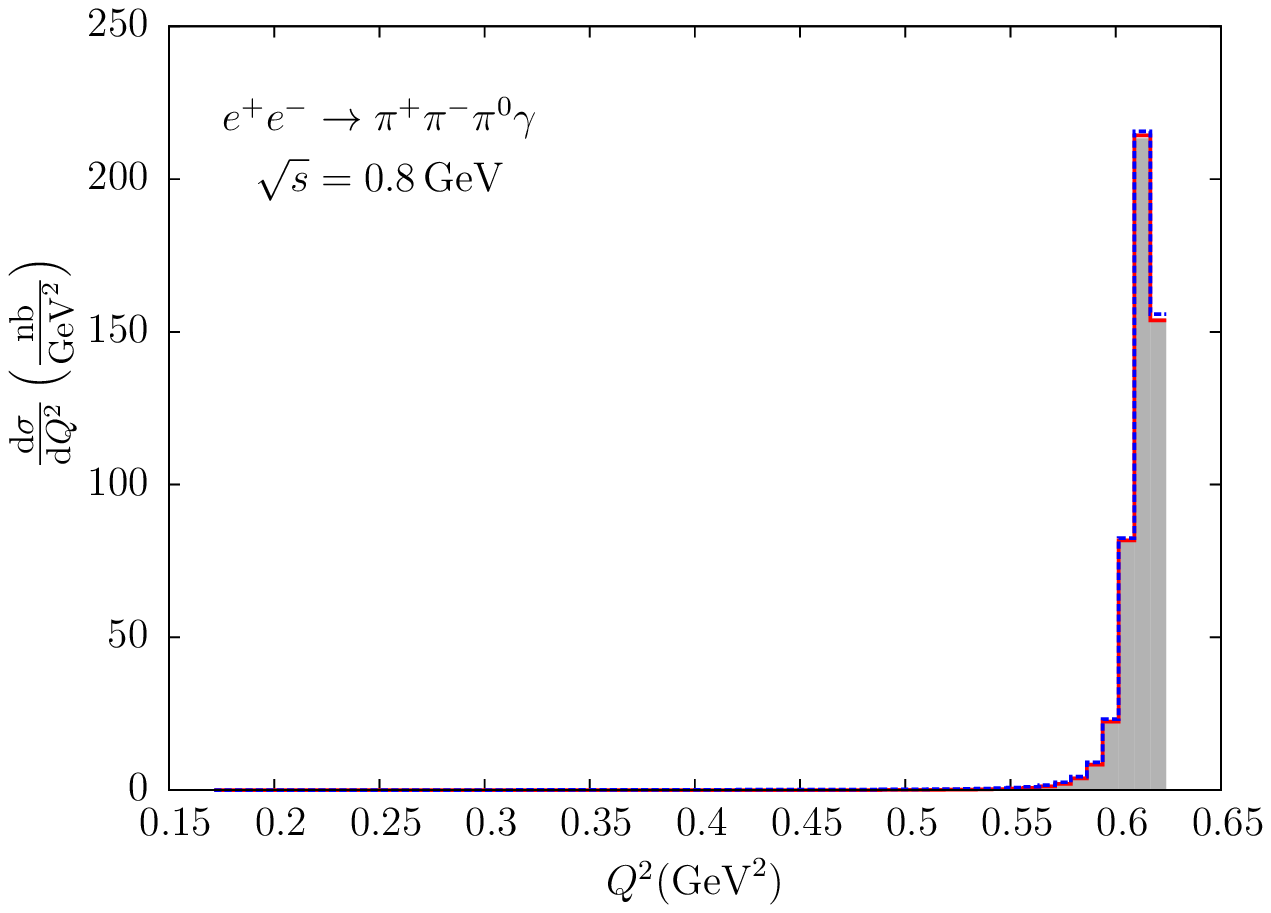,  width=55mm, height=42mm}
\epsfig{file=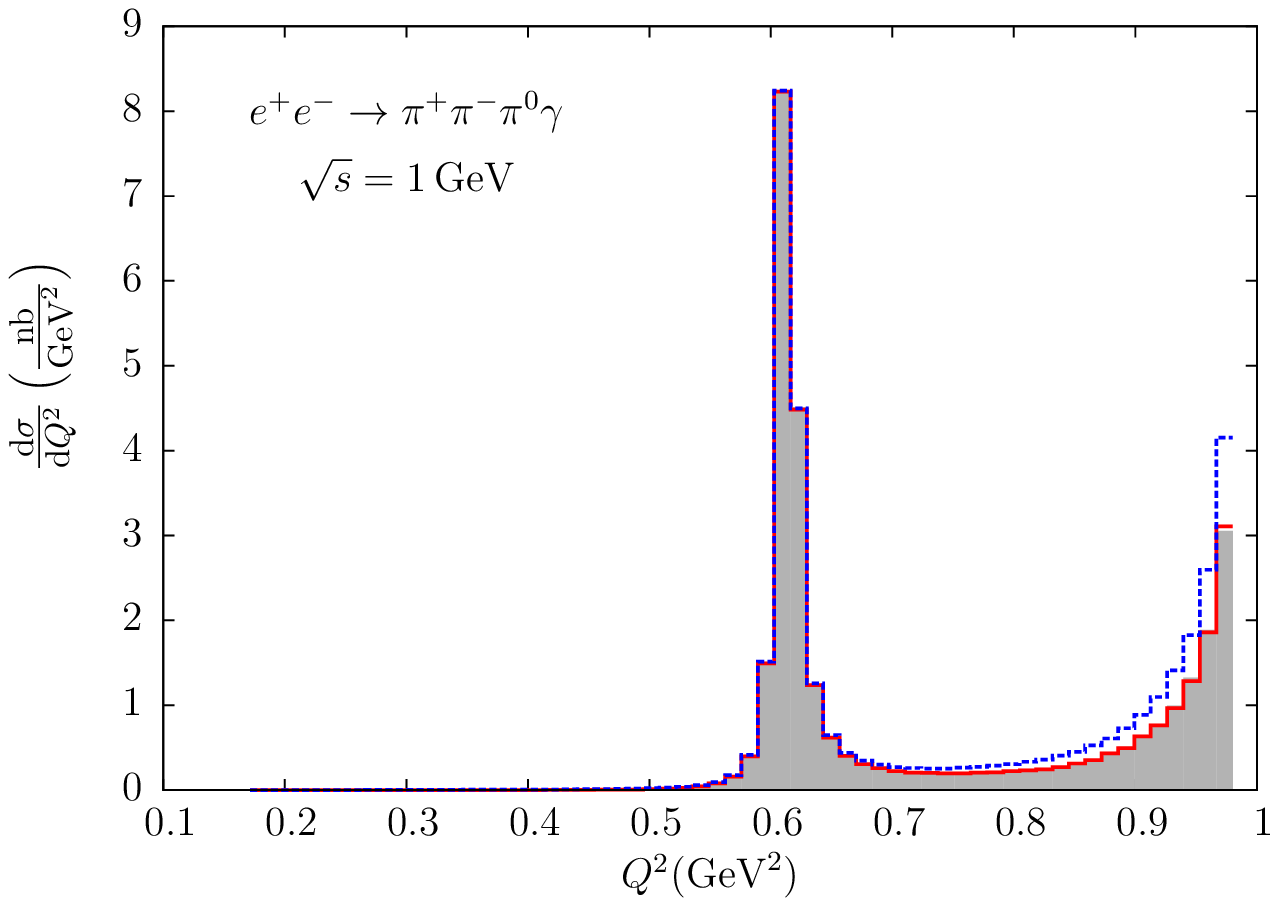,  width=55mm, height=42mm}
\epsfig{file=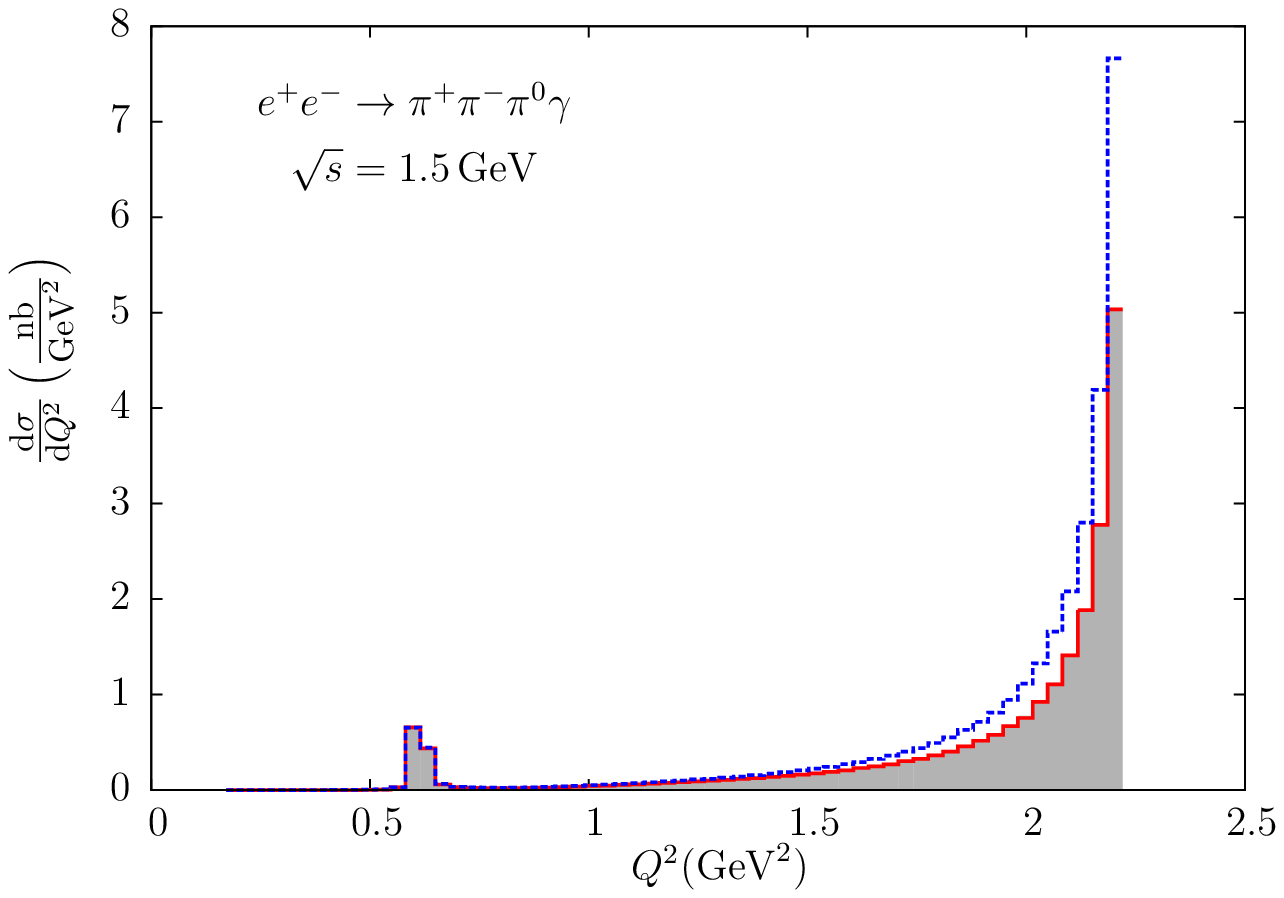,  width=55mm, height=42mm}}
\caption{\small The differential cross sections of (\ref{pppa}) as functions
of invariant mass of the $\pi^+\pi^-\pi^0$-system. Legend as in Fig.~\ref{pp_dQ2}.}
\label{ppp_dQ2}
\end{figure}

\begin{figure}[htb]
\centerline{
\epsfig{file=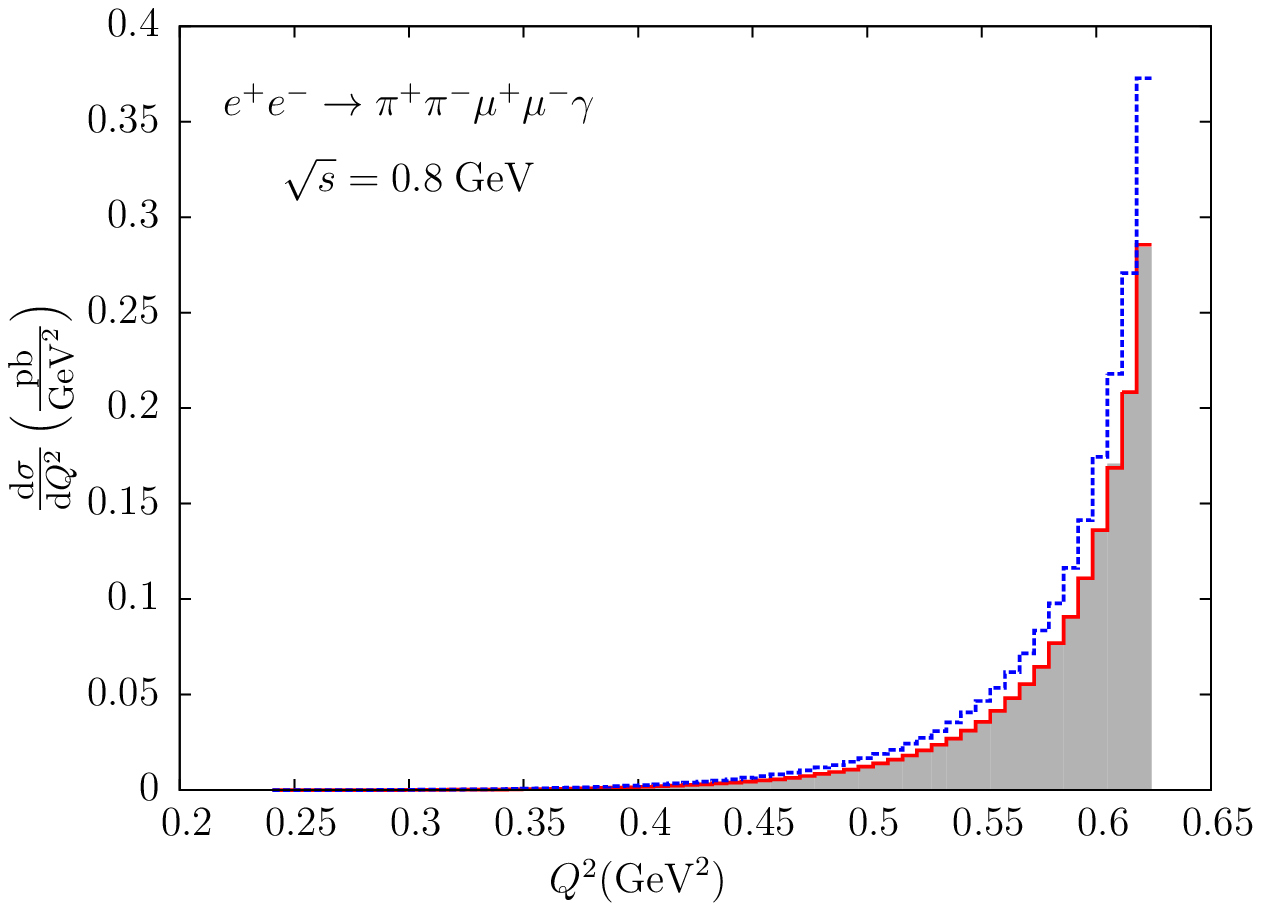,  width=55mm, height=42mm}
\epsfig{file=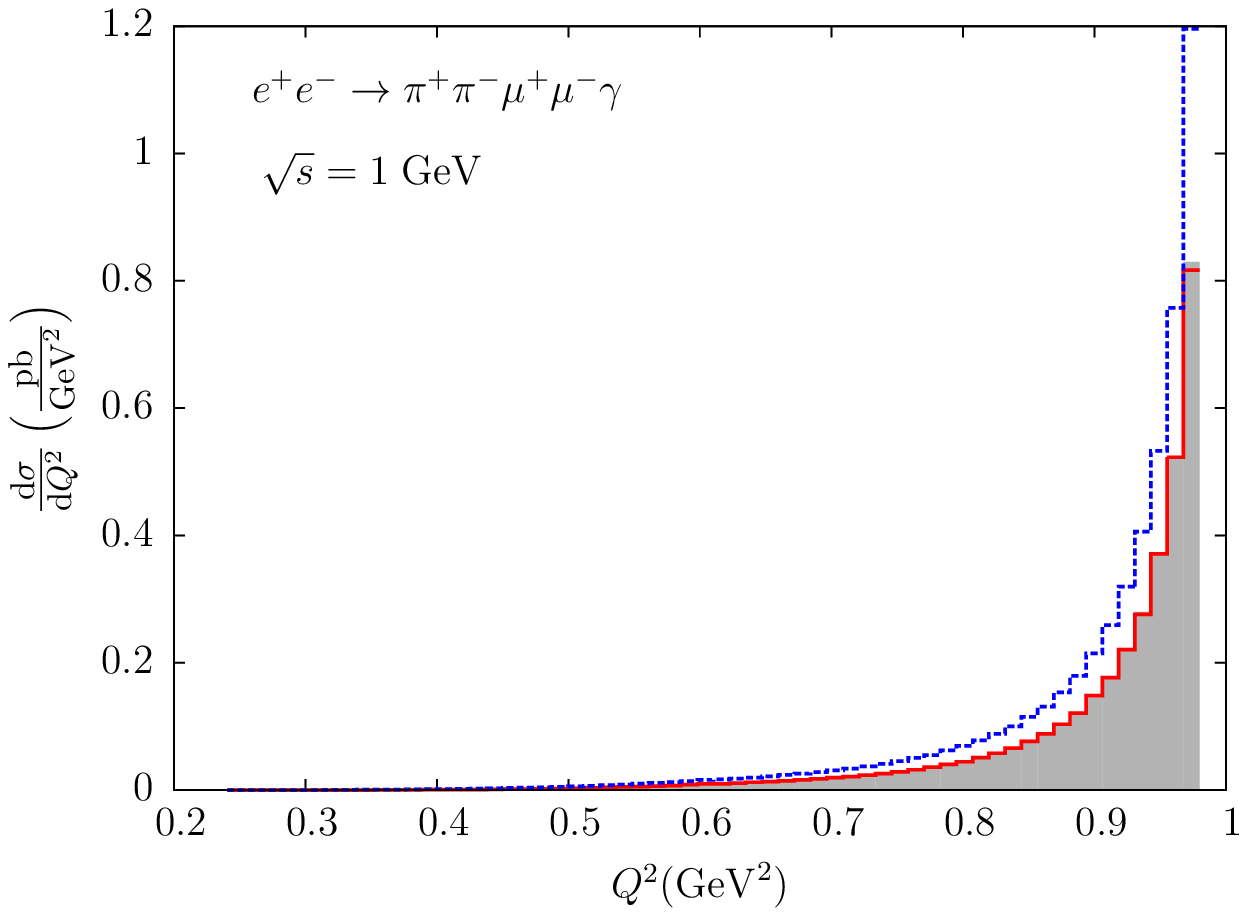,  width=55mm, height=42mm}
\epsfig{file=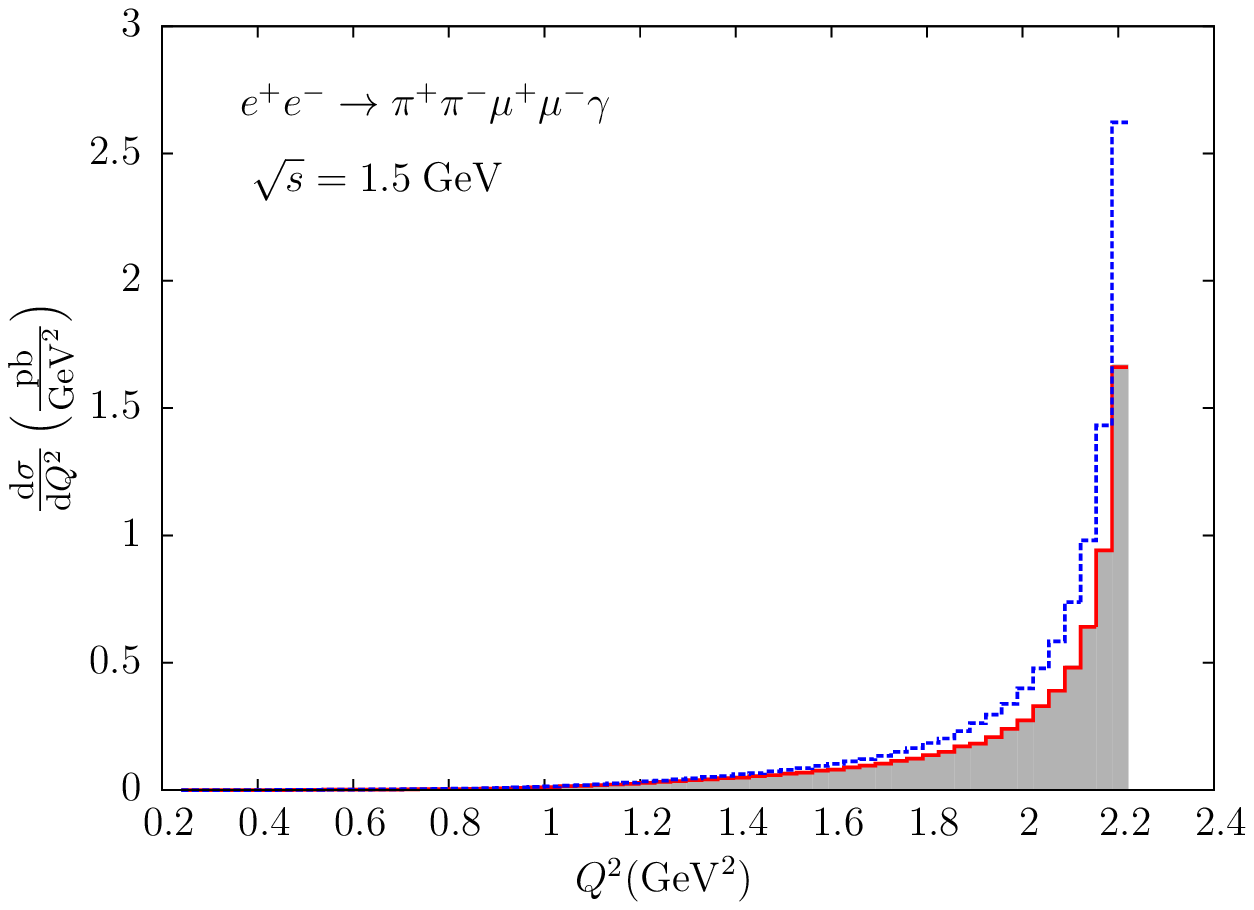,  width=55mm, height=42mm}}
\caption{\small The differential cross sections of 
(\ref{ppmma}) as functions of invariant mass of the $\pi^+\pi^-\mu^+\mu^-$-system.
Legend as in Fig.~\ref{pp_dQ2}.}
\label{ppmm_dQ2}
\end{figure}

\begin{figure}[htb]
\centerline{
\epsfig{file=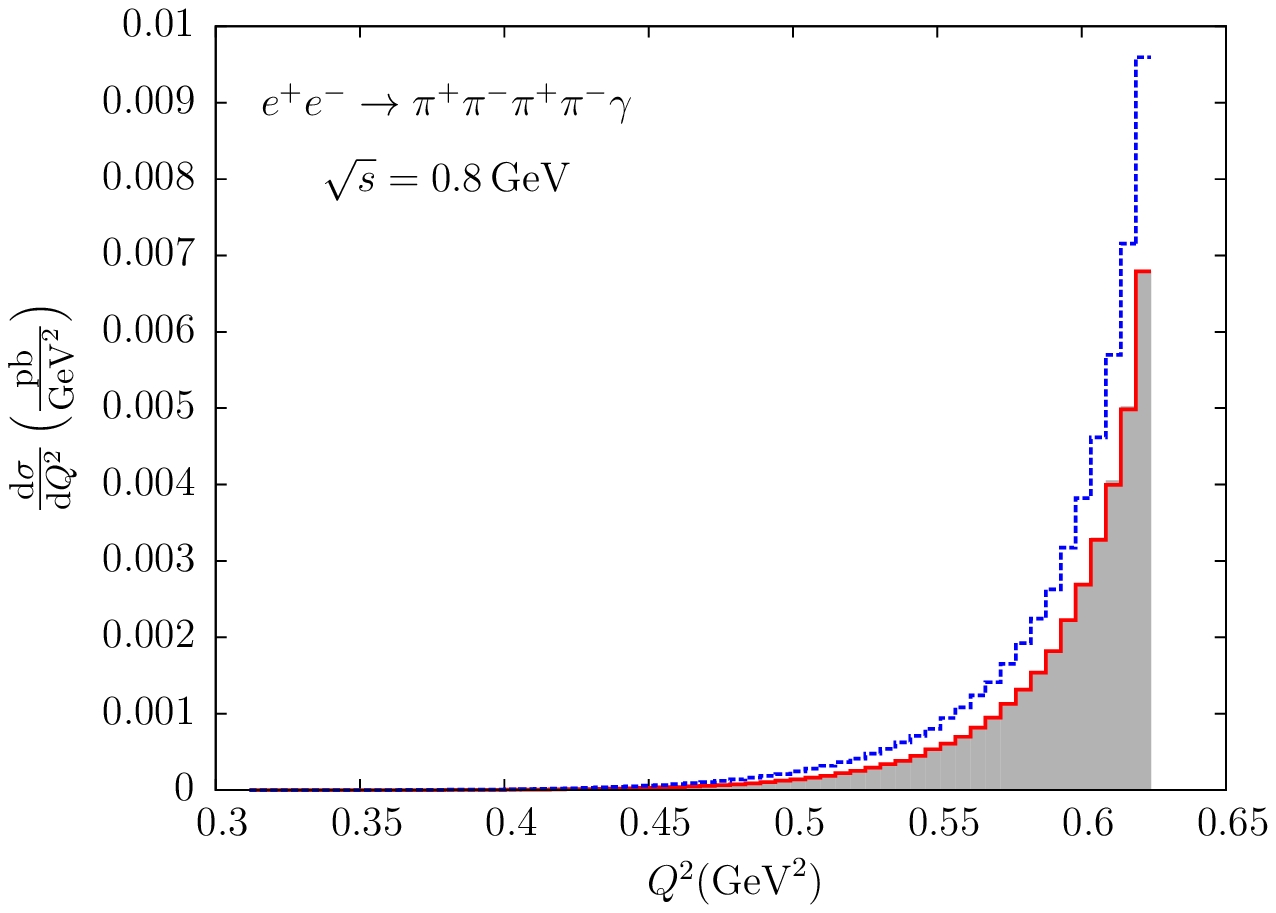,  width=55mm, height=42mm}
\epsfig{file=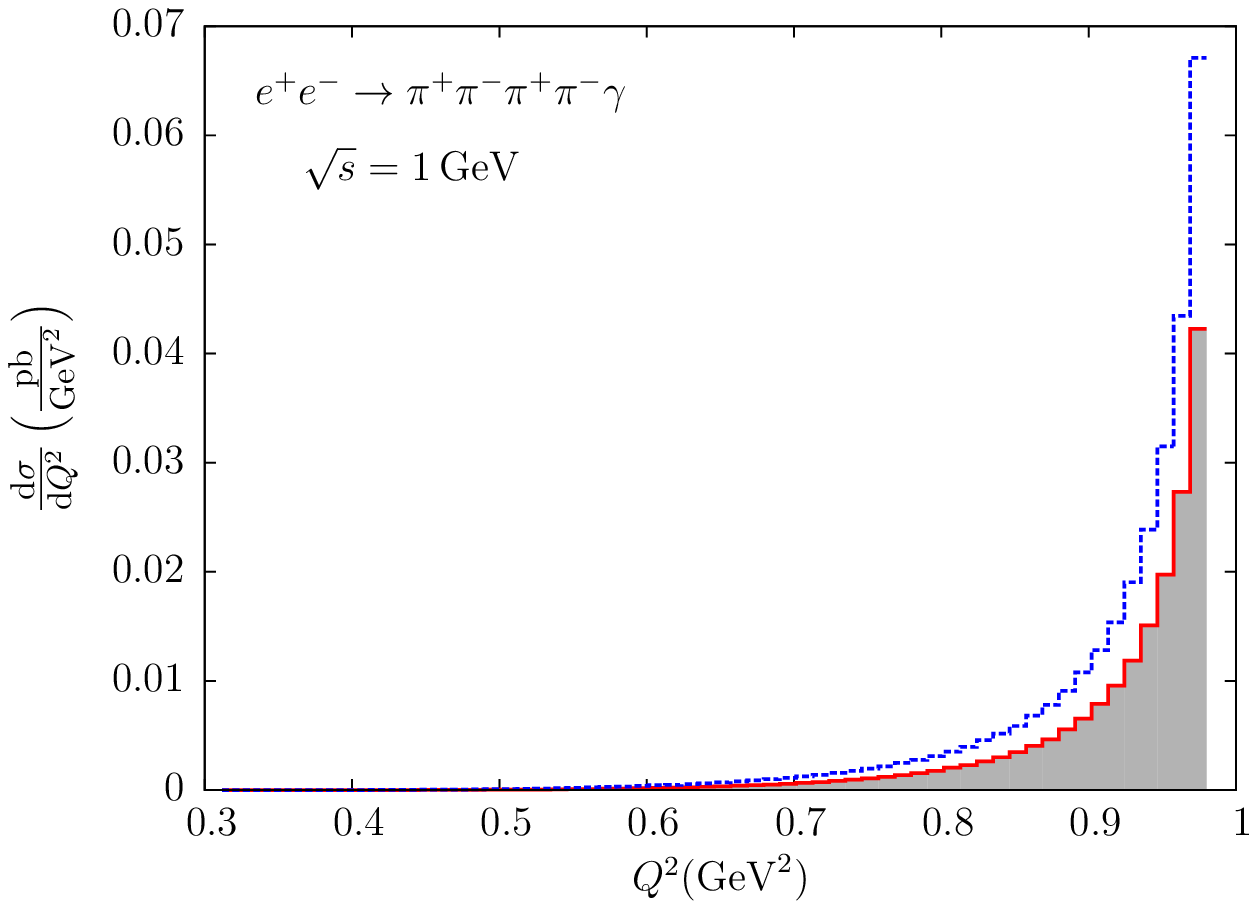,  width=55mm, height=42mm}
\epsfig{file=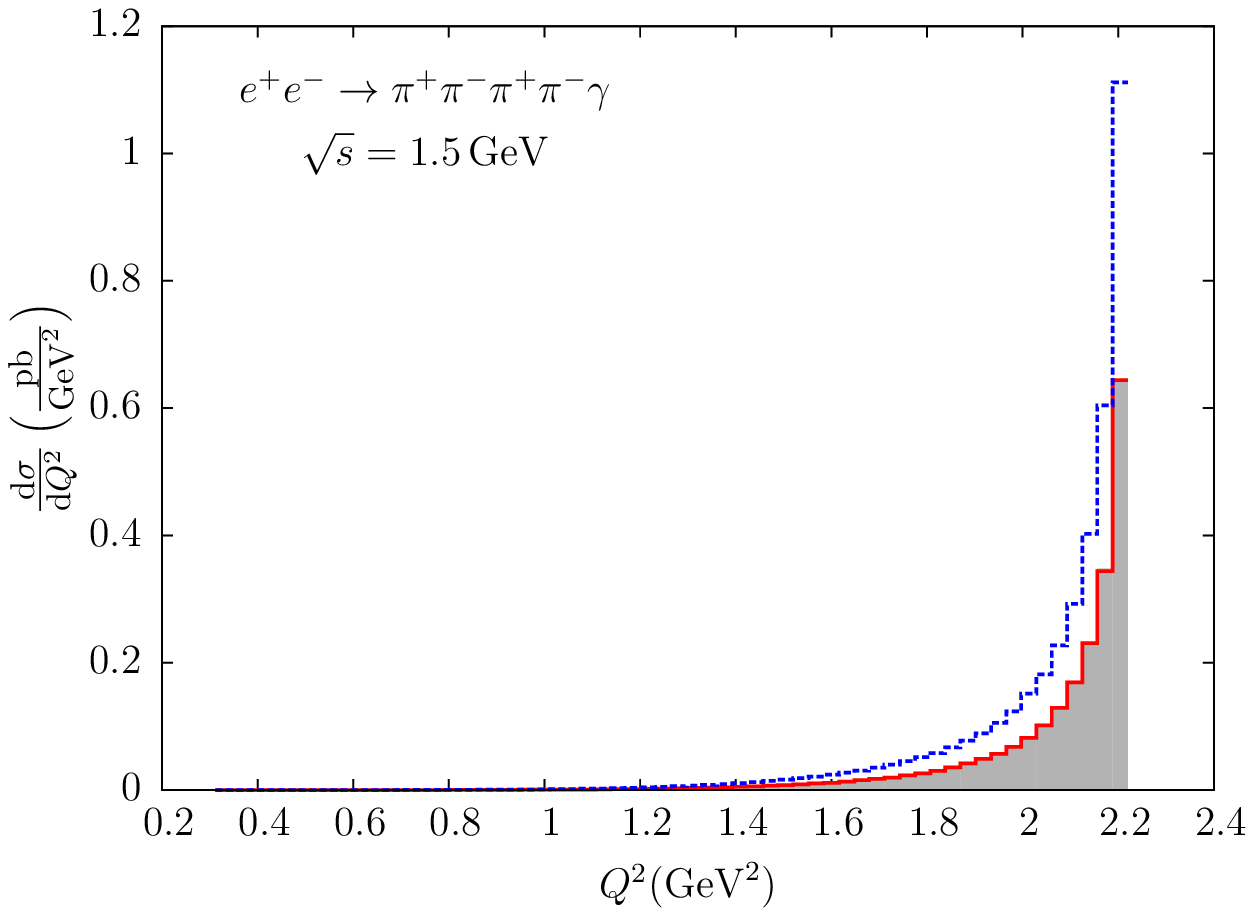,  width=55mm, height=42mm}}
\caption{\small The differential cross sections of (\ref{ppppa})
as functions of invariant mass of the $\pi^+\pi^-\pi^+\pi^-$-system.
Legend as in Fig.~\ref{pp_dQ2}}
\label{pppp_dQ2}
\end{figure}

The differential cross sections of processes (\ref{pppa}), (\ref{ppmma}) and
(\ref{ppppa}) corresponding to those plotted in Fig.~\ref{pp_dQ2} are shown,
respectively, in Figs. \ref{ppp_dQ2}, \ref{ppmm_dQ2} and \ref{pppp_dQ2} as
functions of the invariant mass of the $\pi^+\pi^-\pi^0$-,
$\pi^+\pi^-\mu^+\mu^-$-, or $\pi^+\pi^-\pi^+\pi^-$-system. As can be seen in 
Fig.~\ref{pppp_dQ2}, the FSR effect in the four pion cross section at the 
c.m.s. energies presented is quite substantial. Although we do not present
here the cross sections at energies relevant for the radiative return analysis
of that channel at $B$ factories, which would go well beyond the scope of
the present work, we expect that the FSR effects
are much smaller relative to the ISR for higher energies, as e.g. the energy
of $\Upsilon(4S)$ meson. 

\begin{figure}[htb]
\centerline{
\epsfig{file=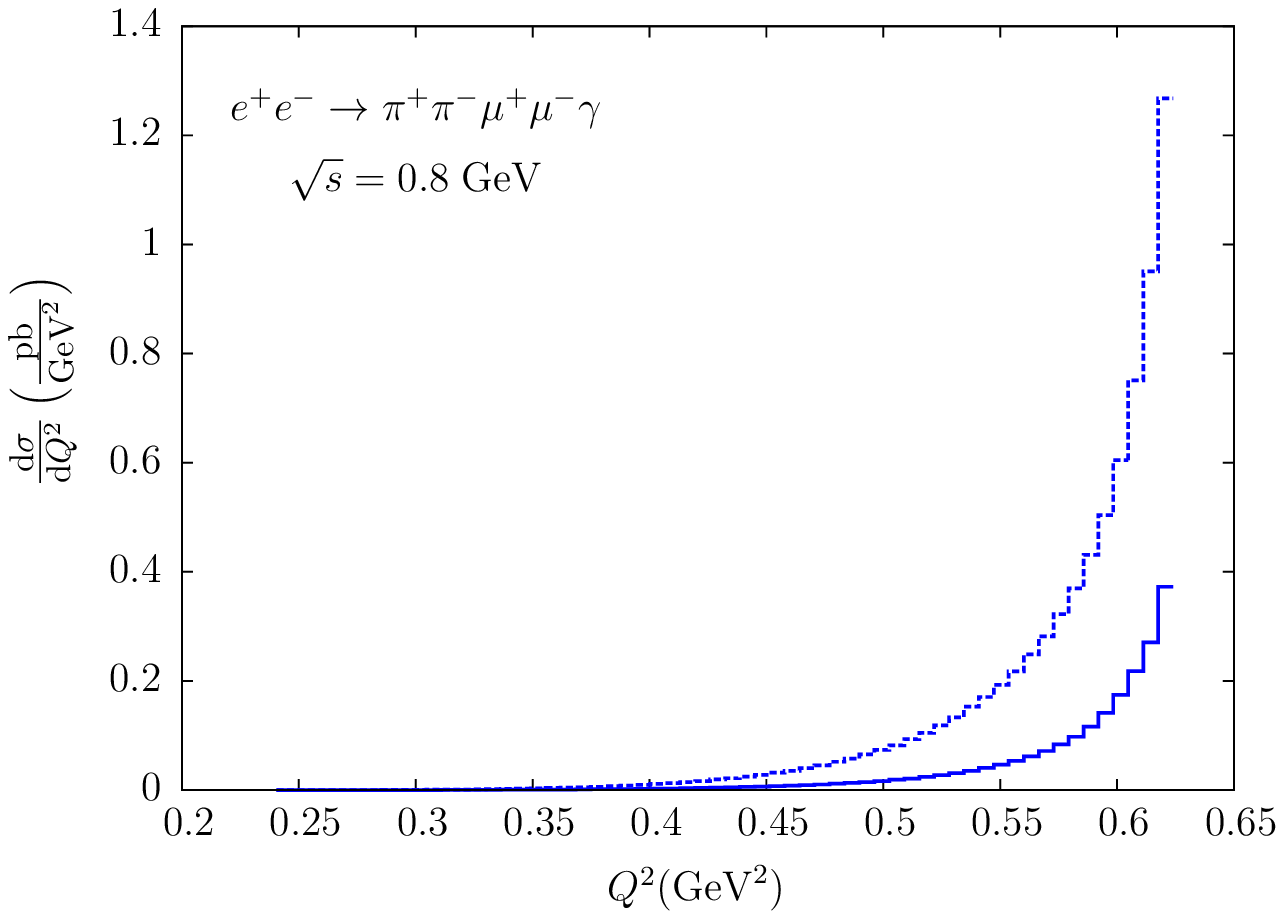,  width=55mm, height=42mm}
\epsfig{file=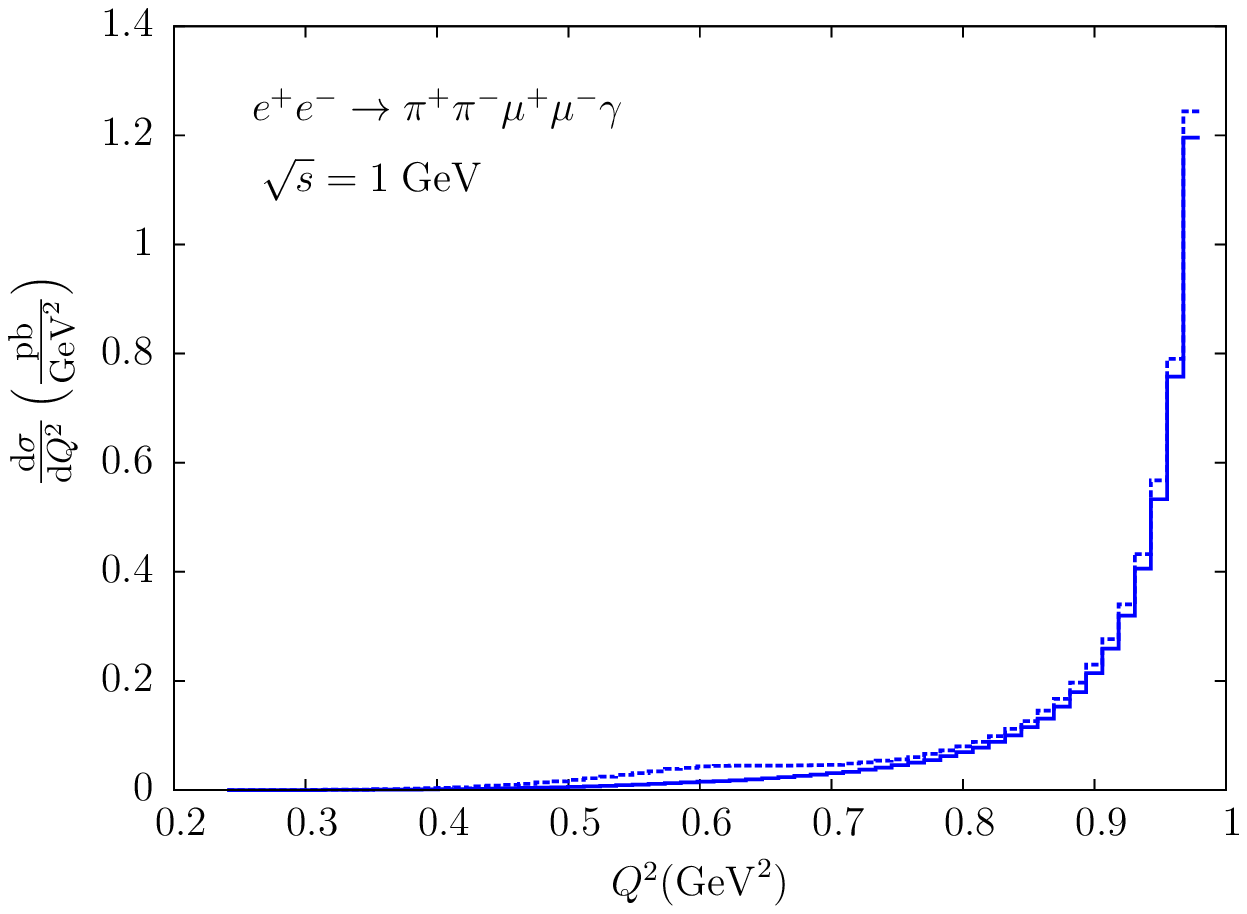,  width=55mm, height=42mm}
\epsfig{file=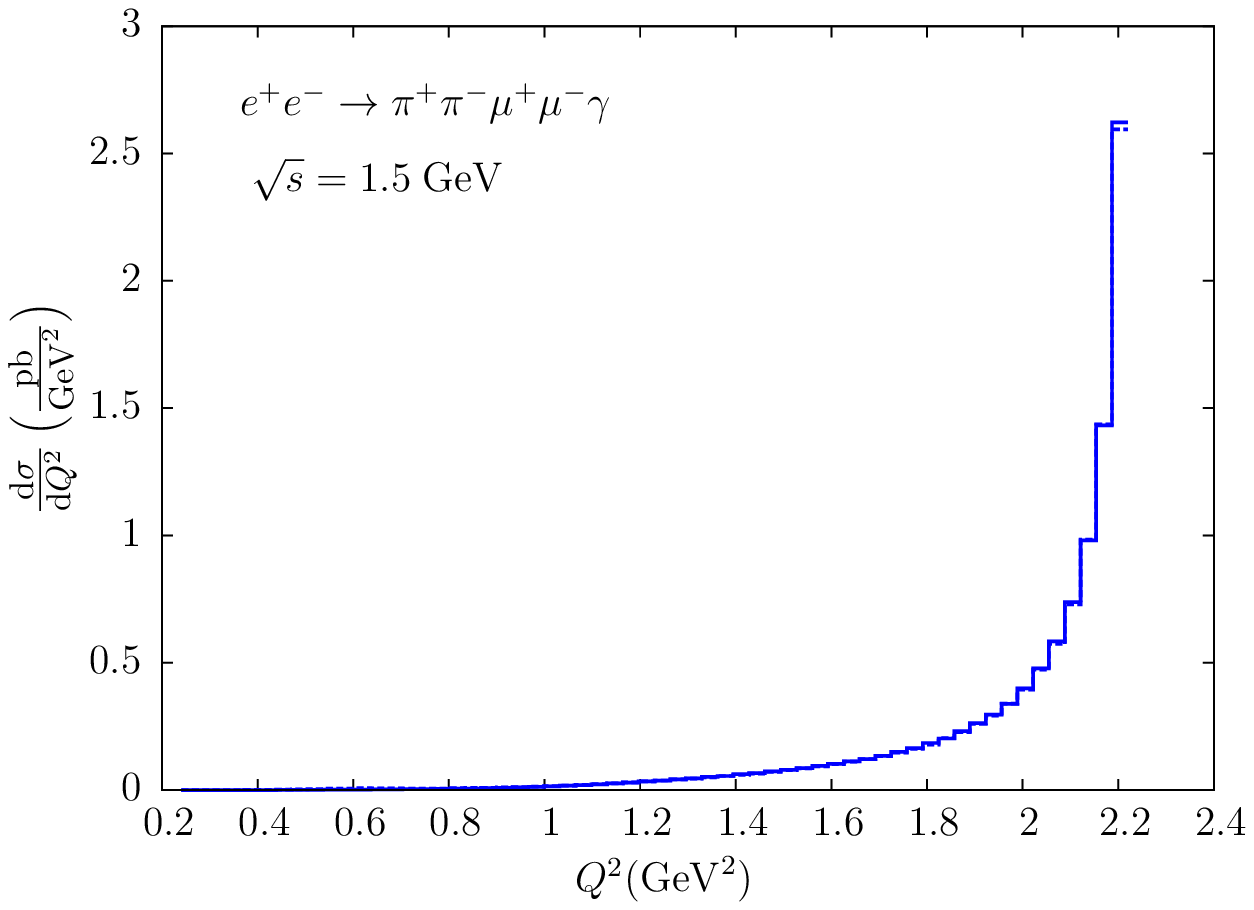,  width=55mm, height=42mm}}
\caption{\small The differential cross sections of (\ref{ppmma}) as functions
  of invariant mass of the $\pi^+\pi^-\mu^+\mu^-$-system computed in a model with
  the charged pion form factor ({\em solid lines}) and in the HLS model with fixed
  couplings ({\em dotted lines}).}
\label{ppmma_ff_vs_diags}
\end{figure}

In Fig.~\ref{ppmma_ff_vs_diags}, we make the same comparison for the cross
sections of process (\ref{ppmma}) as was made in Fig.~\ref{ppa_ff_vs_diags} for
process (\ref{ppa}). This time {\tt carlomat\_3.1} generates 627 Feynman diagrams
in the HLS model with the fixed couplings.

\begin{figure}[htb]
\centerline{
\epsfig{file=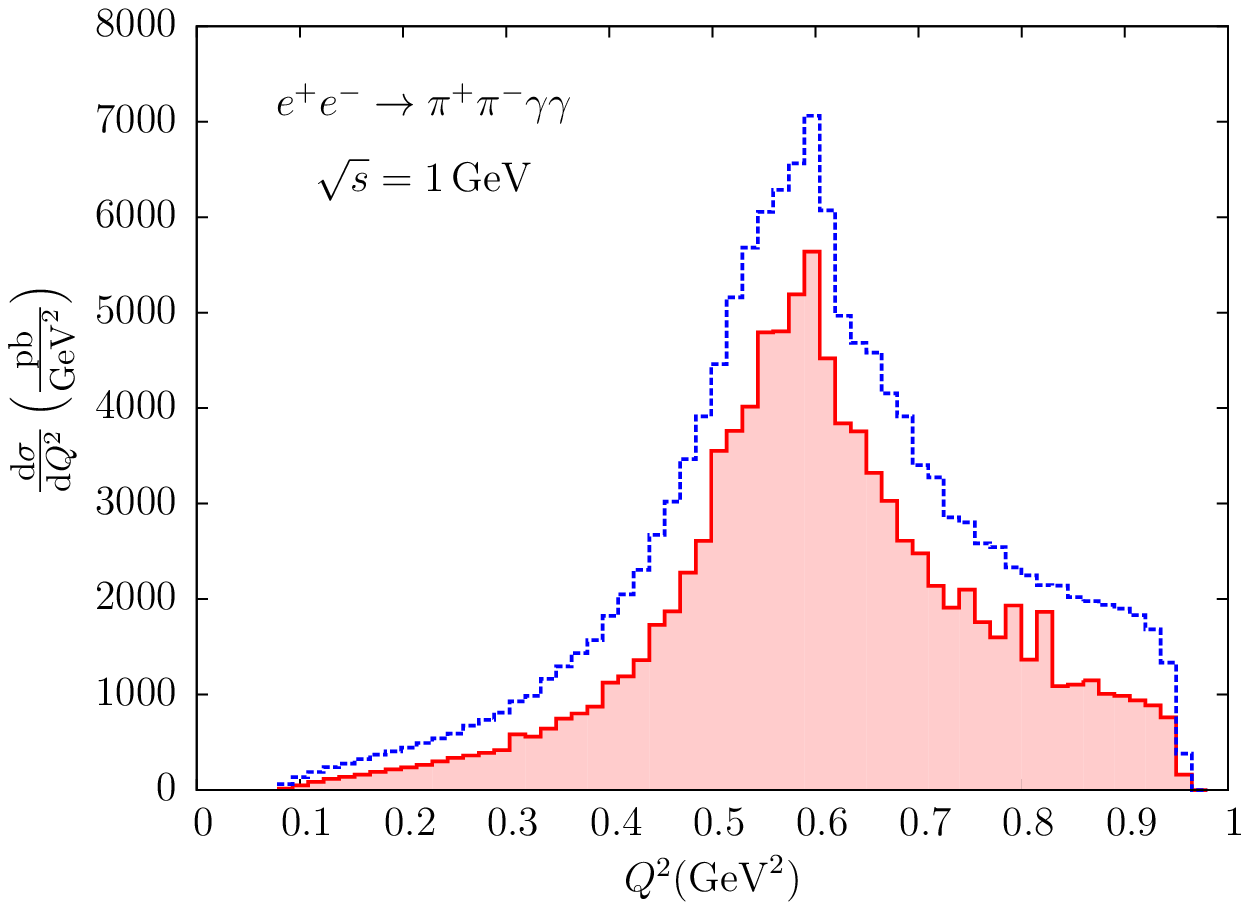,  width=55mm, height=42mm}
\epsfig{file=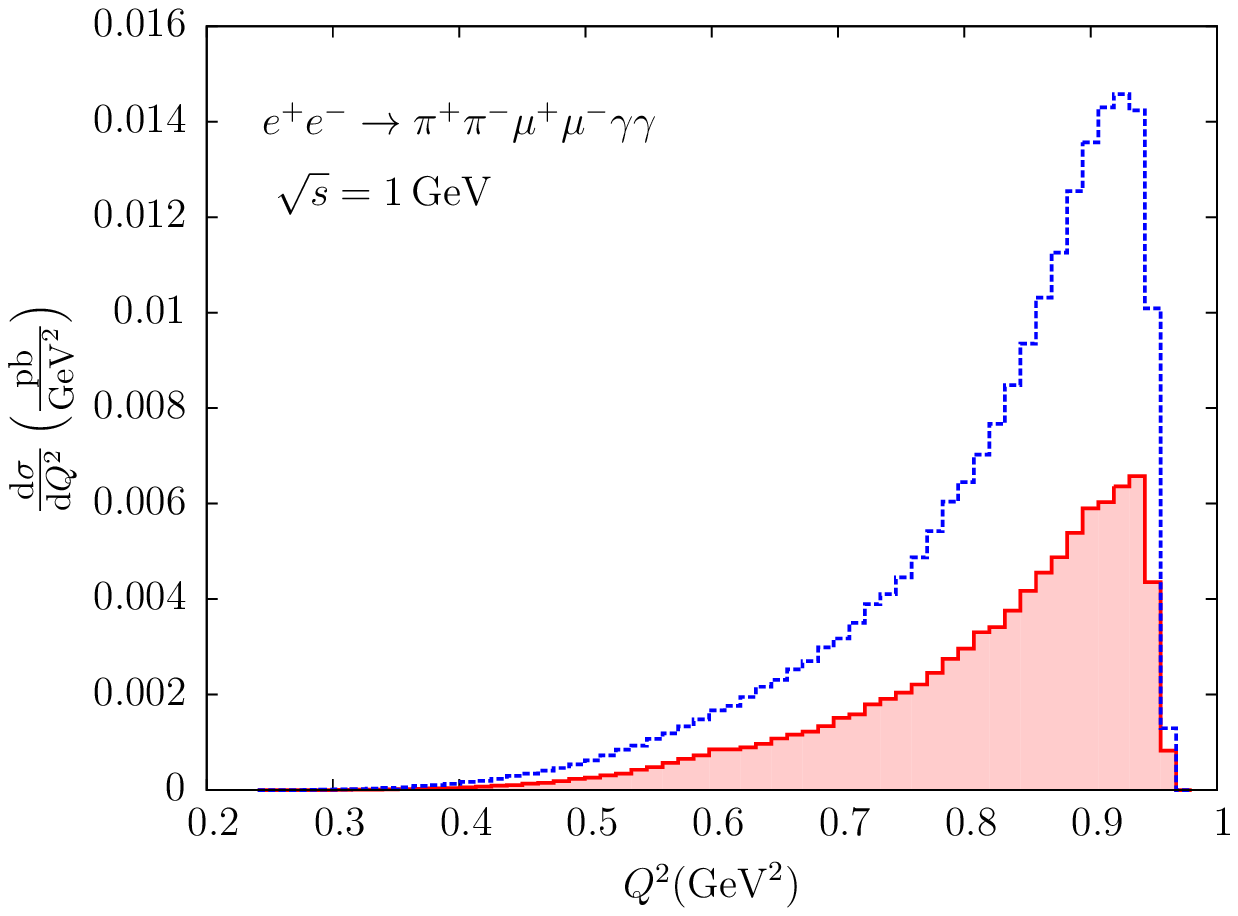,  width=55mm, height=42mm}
\epsfig{file=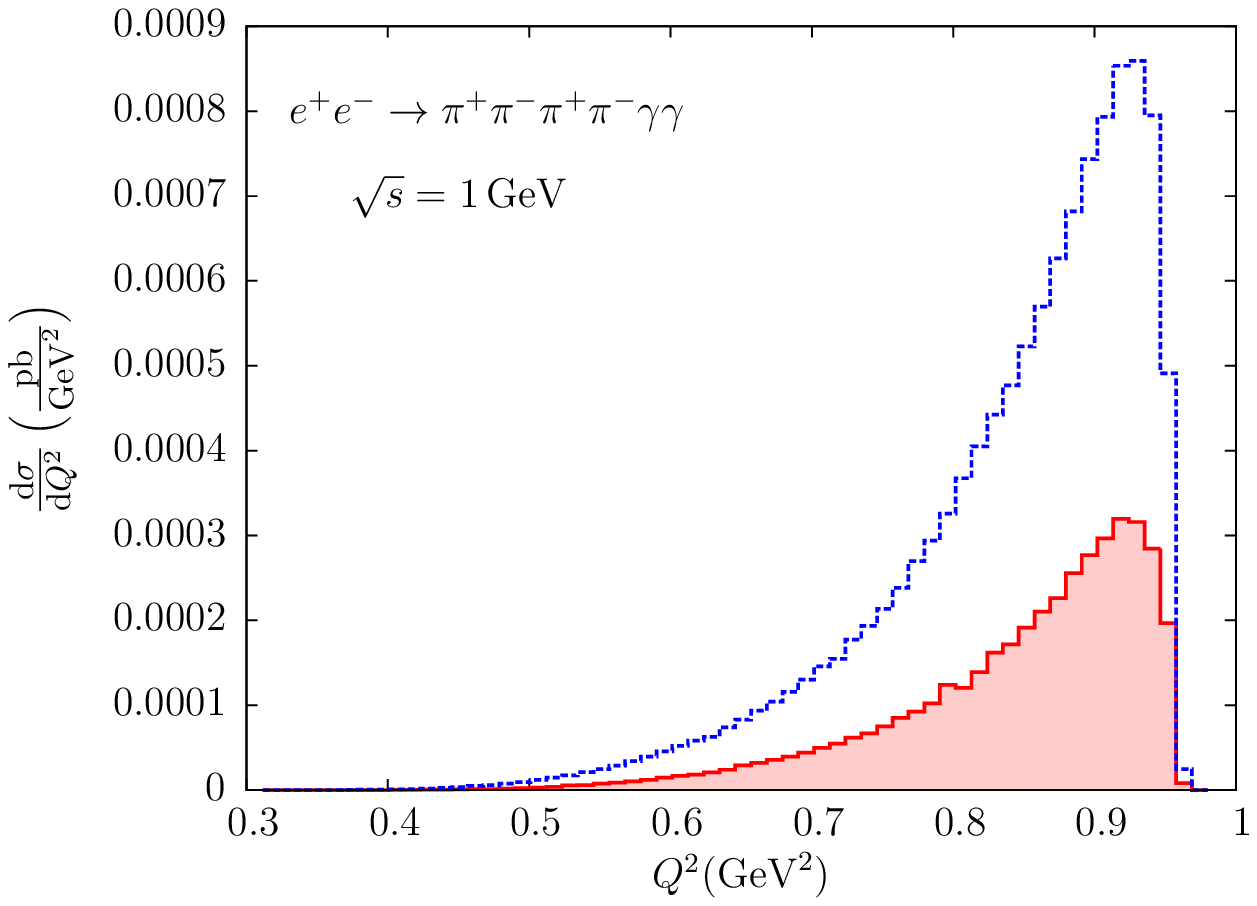,  width=55mm, height=42mm}}
\caption{\small The differential cross sections of (\ref{ppaa}), (\ref{ppmmaa}) 
  and (\ref{ppppaa}) at $\sqrt{s}=1$~GeV as functions of invariant mass of the
  $\pi^+\pi^-$-, $\pi^+\pi^-\mu^+\mu^-$ and $\pi^+\pi^-\pi^+\pi^-$-systems,
  respectively.}
\label{aa_dQ2}
\end{figure}

To illustrate the effect of the FSR in processes with two photons,
the ISR (shaded histograms) and full LO (dashed lines) differential cross sections 
of processes (\ref{ppaa}), (\ref{ppmmaa}) and (\ref{ppppaa}) at $\sqrt{s}=1$~GeV 
are plotted in Fig.~\ref{aa_dQ2} as functions of invariant mass of the
$\pi^+\pi^-$-, $\pi^+\pi^-\mu^+\mu^-$ and $\pi^+\pi^-\pi^+\pi^-$-systems,
respectively. We see that this time the FSR effect is much bigger.

\section{Summary}

We have implemented the electromagnetic charged pion form factor in {\tt carlomat\_3.1},
a new version of a multipurpose program {\tt carlomat} that allows one to generate 
automatically the MC programs dedicated to the description of, among others, the processes 
$\epm\to{\rm hadrons}$ at low centre-of-mass energies. We have illustrated possible 
applications of the program by considering a photon radiation off the initial and final 
state particles for a few potentially interesting processes involving charged pion pairs.
We have also discussed some problems related to the $U(1)$ electromagnetic gauge 
invariance that may arise if the momentum transfer dependence is introduced
in the couplings of the HLS model or a set of the couplings implemented in the program
is incomplete.

\end{document}